\documentclass[twocolumn,english,english,preprintnumbers,amsmath,amssymb,prb,superscriptaddress]{revtex4}
\usepackage[T1]{fontenc}
\usepackage[latin9]{inputenc}
\usepackage{amsmath}
\usepackage{amssymb}
\usepackage{esint}
\usepackage[normalem]{ulem}

\makeatletter
\@ifundefined{textcolor}{}
{%
 \definecolor{BLACK}{gray}{0}
 \definecolor{WHITE}{gray}{1}
 \definecolor{RED}{rgb}{1,0,0}
 \definecolor{GREEN}{rgb}{0,1,0}
 \definecolor{BLUE}{rgb}{0,0,1}
 \definecolor{CYAN}{cmyk}{1,0,0,0}
 \definecolor{MAGENTA}{cmyk}{0,1,0,0}
 \definecolor{YELLOW}{cmyk}{0,0,1,0}
 }

\usepackage{graphics}\usepackage{epstopdf}\usepackage{epsfig}\@ifundefined{definecolor}{\usepackage{color}}{}
\usepackage{amsfonts}\usepackage{bm}\usepackage{amscd}

\input{tcilatex}

\usepackage{babel}

\usepackage{babel}

\usepackage{babel}

\usepackage{babel}

\usepackage{babel}

\makeatother

\usepackage{babel}
\begin{document}

\title{Theory of wavepacket transport under narrow gaps and spatial textures: non-adiabaticity and semiclassicality}

\author{Matisse Wei-Yuan Tu}
\email{kerustemiro@gmail.com}
\affiliation{Department of Physics, The University of Hong Kong, China}
\affiliation{HKU-UCAS Joint Institute of Theoretical and Computational Physics at Hong Kong, China}
\author{Ci Li}
\affiliation{Department of Physics, The University of Hong Kong, China}
\affiliation{HKU-UCAS Joint Institute of Theoretical and Computational Physics at Hong Kong, China}
\author{Wang Yao}
\affiliation{Department of Physics, The University of Hong Kong, China}
\affiliation{HKU-UCAS Joint Institute of Theoretical and Computational Physics at Hong Kong, China}

\begin{abstract}
We generalise the celebrated semiclassical wavepacket approach from the adiabatic to the non-adiabatic regime. A unified description covering both of these regimes is particularly desired for systems with spatially varying band structures where band gaps of various sizes are simultaneously present, e.g. in moir\'{e} patterns. For a single wavepacket, alternative to the previous derivation by Lagrangian variational approach, we show that the same semiclassical equations of motion can be obtained by introducing a spatial-texture-induced force operator similar to the Ehrenfest theorem. For semiclassically computing the current, the ensemble of wavepackets based on adiabatic dynamics is shown to well correspond to a phase-space fluid for which the fluid's mass and velocity are two distinguishable properties. This distinction is not inherited to the ensemble of wavepackets with the non-adiabatic dynamics. We extend the adiabatic kinetic theory
to the non-adiabatic regime by taking into account decoherence, whose joint action with electric field favours certain form of inter-band coherence. The steady-state density matrix as a function of the phase-space variables is then phenomenologically obtained for calculating the transport current. The result, applicable with a finite electric field, expectedly reproduces the known adiabatic limit by taking the electric field to be infinitesimal, and therefore attains a unified description from the adiabatic to the non-adiabatic situations.
\end{abstract}
\maketitle

\section{Introduction}

The so-called semiclassical wavepacket approach has been successfully
applied to study the motion of electrons in crystals.\cite{Ashcroft76book,Xiao101959}
Most notably in the semiclassical equations of motion (SC-EOM) for the wavepacket,
anomalous velocities in terms of the Berry curvatures play a key role in understanding the steady transport properties for a number of phenomena, including the anomalous Hall effect,\cite{Karplus541154,Luttinger58739,Adams59286,Jungwirth02207208,Nagaosa101539}
spin Hall effect,\cite{Murakami031348,Sinova04126603}
the valley Hall effect\cite{Xiao07236809,Xiao12196802} as well as chiral anomaly in Weyl metals.\cite{Son13104412} The non-sympletic structure of the SC-EOM\cite{Xiao05137204,Chang08193202} also raises interests in its real-time dynamics in the context of Dirac semimetals \cite{Gorbar18045203}
and diffusive processes.~\cite{Olson07035114,Misaki18075122}
The potential utility of semiclassical wavepacket approach merits further attention.

%Since only those few bands whose energies are near the fermi energy are relevant for low-energy properties, the active manifold can be either the conduction or the valence bands if they are separated by a sufficiently large gap. Otherwise, if the electron and hole bands are closely spaced, with respect to the applied field, or even degenerate, then a single active manifold consisting of both sets of bands is required.

\begin{figure}[h]
 \includegraphics[width=9.8cm,height=6.8cm]{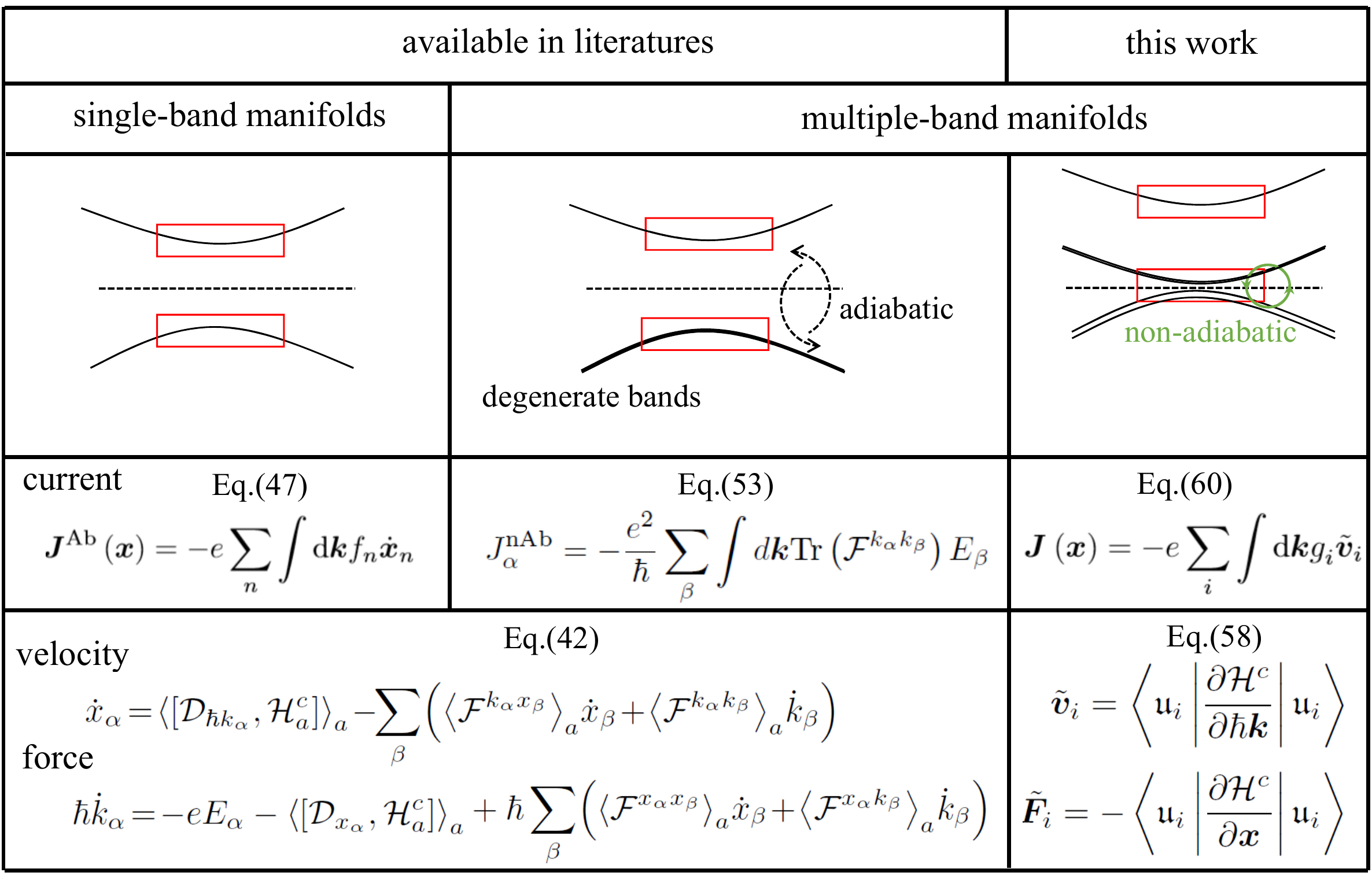}
\caption{The bands are grouped into manifolds, each indicated by a red box. Inter-manifold energy spacing is much larger than the applied electric field.
The two left columns are the scenarios where each manifold either has a single band only, or consists of fully degenerate bands. The corresponding currents are given respectively by Eq.~(\ref{Bltzeq-2}) and Eq.~(\ref{anom-crnt-Ad-nAB-0}), which directly make use of centre-of-mass SC-EOM, Eq.~(\ref{phsvelo-sep-2}) (and its single-band reduction, Eq.~(\ref{phsvelo-sep-2-AB})). In the right-most column, the manifold enclosing the Fermi energy (dashed line) consists of several non-degenerate bands with small energy spacing/gap. Non-adiabatic generalisation takes into account decoherence, whose joint action with driving electric field favours certain form of inter-band coherence, which affects centre-of-mass variables as given by Eq.~(\ref{veloeq-0}) and results in a current given by Eq.~(\ref{Bltzeq-4}). See Sec.~\ref{semiCcrnt-1} for detailed discussions.}
\label{schematics}
\end{figure}

In general, given the strength of the applied electric field, the bands of an electronic material can be grouped into manifolds, see Fig.~\ref{schematics}. Those bands whose spacings are small in comparison to the energy scale associated with the external field are grouped into the same manifold. By definition, inter-manifold energy spacing is much larger than the applied field. When one focuses on the motion of a wavepacket within a particular manifold, the effects of other manifolds are manifested by the Berry curvatures in the SC-EOM.~\cite{Xiao101959} We call the manifold of focus as the active manifold and the bands within that manifold as active bands.
The already known semiclassical wavepacket approaches mainly deal with two situations in the active manifold. The first is that the active manifold has only one band with Abelian Berry curvatures.~\cite{Chang951348,Chang967010,Sundaram9914915} The second is that it consists of several degenerate bands with non-Abelian Berry curvatures.~\cite{Culcer05085110,Shindou05399} For both cases, given the large gaps between different manifolds, the effect of an electric field whose strength is small in comparison to these gaps can be well captured by the first order adiabatic approximation, which accounts the electric field to the first order. However, in addition to these two situations, there is a third situation where the active manifold contains several non-degenerate bands. For this third case, although the effect of the electric field on inter-manifold transition can be accounted by the same approximation, its effect on the intra-manifold transitions, due to the relatively small energy separations, should be taken into account to all orders and is expected to induce non-adiabatic dynamics. Henceforth, throughout this article, we term this third case as non-adiabatic while the two former cases as adiabatic. The purpose of the present work is to generalise the semiclassical wavepacket approach developed in the series of studies~\cite{Chang951348,Chang967010,Sundaram9914915,Culcer05085110} and reviewed in Ref.~[\onlinecite{Xiao101959}] for the adiabatic situation to the non-adiabatic one. This is motivated by the following realistic considerations.

(i): Berry curvature
is inversely proportional to the square of gap, and therefore features hot spot at the small gaps typically arising from band anti-crossings. Examples include the gapped graphene,~\cite{Hunt131427,Woods14451} and the recently discovered 2D $\text{MnBi}_{2}\text{Te}_{4}$
  (with gap sizes of tens of meV).~\cite{Zhang19206401,Li19121103,Li19eaaw5685,Liu2005733} Electron transport under finite  electric field raises the need of addressing the non-adiabaticity.

(ii): The wavepacket approach can directly address spatially varying band structures, in addition to momentum textures, exemplified by its application to systems with deformation potentials~\cite{Sundaram9914915} and magnetic textures.\cite{Yang09067201,Everschor-Sitte14172602} Spatially varying band structures are also relevant in long-period Moir\'{e} superlattices (with periods much larger than the lattice constant) of high current interest. Experimentally, the location dependence of gap sizes has already been observed.~\cite{Zhang17e1601459,Pan181849} However, theoretical understanding of the electronic structures is mainly limited to the Moir\'{e} mini-bands,~\cite{Bistritzer1112233,Kindermann12115415,Wallbank15359} which treats the long-period Moir\'{e} pattern from the perspective of global band structure and henceforth does not address explicitly the spatial textures. The wavepacket approach is on one hand complementary to the mini-band picture and on the other hand beneficial when the Moir\'{e} pattern is non-periodic, as found in most experimental realities. The non-adiabatic effects are sometimes inevitable due to Moir\'{e} spatial textures.  Gapped graphenes on hexagonal boron nitrides~\cite{Hunt131427,Woods14451} is such an example, where infinitesimal local gap exists due to sign reversal of the gap as function of location.

Generalisation of the semiclassical wavepacket approach developed in the adiabatic regime to the non-adiabatic regime is not straightforward. This can be seen from the two primary steps in constructing a semiclassical wavepacket theory for studying electron transport.\cite{Ashcroft76book,Xiao101959} In the first step, one obtains the SC-EOM for the centre-of-mass of a single wavepacket. For the second step, one considers an electron gas as ensemble of wavepackets and computes the current $\boldsymbol{J}$ by,~\cite{Ashcroft76book,Xiao101959,Chang951348,Chang967010,Shindou05399}
\begin{equation}
\boldsymbol{J}=-e\int\text{d}\boldsymbol{k}f\left(\boldsymbol{k}\right)\dot{\boldsymbol{x}},\label{sclJ1}
\end{equation} where $\dot{\boldsymbol{x}}$ is the velocity of the wavepacket obtained from the SC-EOM in the first step, and $f\left(\boldsymbol{k}\right)=f_{0}\left(\boldsymbol{k}\right)+\delta f\left(\boldsymbol{k}\right)$ is the carrier distribution
function in which $f_{0}\left(\boldsymbol{k}\right)$ is the equilibrium part and $\delta f$ is the deviation from equilibrium. For the adiabatic cases, where the active manifold has only one band or degenerate bands, $f_{0}\left(\boldsymbol{k}\right)$ can be unambiguously assigned by the fermi distribution function evaluated with that particular band energy. However, in the non-adiabatic situation, the electron has inter-band coherence among a number of bands with distinct energies and various occupations within the active manifold. Even just evaluating the equilibrium part of the distribution function raises ambiguity.

Fig.~\ref{schematics} tabulates available formulae as our reference for making generalisation in the first two columns. The third column summarises the generalised results from this work.
In Sec.~\ref{wavpck-dys-1}, we derive the SC-EOM of a wavepacket using a different approach, without repeating the derivation by the variational approach for wavepacket-based Lagragian.~\cite{Chang951348,Chang967010,Sundaram9914915,Culcer05085110,Xiao101959} How the SC-EOM emerges from a more fundamental quantum consideration has long been interested, within both the conventions of solid-state physics~\cite{Slater491592,Luttinger51814,BlountbookChp62,Wannier62645,Zak68686,Chang967010,Sundaram9914915,Culcer05085110,Shindou05399,Shindou08035110,Lapa19121111,Stedman19103007} and mathematical physics.~\cite{Panati03547,Gosselin06651,Bliokh057,Dayi08315204,Bettelheim17415303} Here we start from the full-band space and introduce a force operator similar to the spirit of the Ehrenfest theorem. We will show that with the aid of the Netownian law for the time-changing rate of momentum, SC-EOM straightforwardly arises. In Sec. \ref{wavpck-dys-2}, we discuss how the Berry curvatures appear in the SC-EOM~\cite{Culcer05085110,Chang967010,Sundaram9914915} when one groups the full bands into manifolds and focuses on a particular one. We will see that when the active manifold contains several bands, non-Abelian Berry curvatures are obtained without requiring exact degeneracy within this manifold, i.e. for the situations illustrated in the second and the third columns of Fig.~\ref{schematics}. By definition, the non-Abelian Berry curvature reduces to Abelian one by reducing the number of bands in the active manifold to one. In Sec.~\ref{semiCcrnt-1}, we turn to an ensemble of wavepackets for computing semiclassically the transport current for an electron gas. The gas of electrons is inspected using the the kinetic theory~\cite{Ashcroft76book} and we extend it to include non-adiabatic effects by taking into account decoherence, whose joint action with electric field favours certain form of inter-band coherence. We then show the reproduction of known results by reducing the active manifold to contain either one band or only degenerate bands. A summary is in Sec.\ref{Sec-con}.

\section{Semiclassical wavepacket dynamics}

\label{Sec-Ehrenfest}

In this section, we first concentrate on the dynamics of a single
wavepacket in a system of multiple bands with spatially varying band
structures. We exclusively consider transport driven only by an external
electric field and we assume there is no external magnetic field. Therefore, the gradient correction
to the wavepacket energy which manifests itself as the orbital magnetic
moment coupled with external magnetic field\cite{Sundaram9914915}
is ignored.

\subsection{Full-band dynamics of an electron wavepacket in spatial and
momentum textures}

\label{wavpck-dys-1}

%Real-space textures, as a manifestation of location dependent band structures,
%can be considered by introducing the so-called local Hamiltonian.

Spatially varying band structures can be obtained by
designating to each local region a Hamiltonian with corresponding
periodic potential. This so-called local Hamiltonian can be understood
as the Hamiltonian experienced by an electron localised as a wavepacket
in the corresponding region, with the capacity to encode the information of spatial
textures.\cite{Culcer05085110,Sundaram9914915,Shindou05399} The real-space
coordinate $\boldsymbol{x}_{c}$ of the wavepacket's centre parameterises
and characterises such local Hamiltonian, denoted as $H_{c}\left(\boldsymbol{x}_{c},t\right)$
where the extra time dependence $t$ comes from a time-dependent vector potential.

The local Hamiltonian possesses Bloch states as its eigenstates
denoted by $\left\vert \psi_{n,\boldsymbol{q}}\left(\boldsymbol{x}_{c},t\right)\right\rangle $
with eigenenergies $\varepsilon_{n,\boldsymbol{q}}\left(\boldsymbol{x}_{c},t\right)$
where $n$ is the band index, $\boldsymbol{q}$ is the momentum quantum
number and $\left(\boldsymbol{x}_{c},t\right)$ reminds us of the
parameterisation of the local Hamiltonian. This guarantees that the
corresponding Schr\"{o}dinger equation,
\begin{equation}
H_{c}\left(\boldsymbol{x}_{c},t\right)\left\vert \Phi\left(t\right)\right\rangle =i\hbar\left\vert \dot{\Phi}\left(t\right)\right\rangle ,\label{locSchEq-1}
\end{equation}
 has a solution of the form,
\begin{equation}
\left\vert \Phi\left(t\right)\right\rangle =\sum_{n}\eta_{n}\left(t\right)\left\vert \psi_{n,\boldsymbol{q}_{c}}\left(\boldsymbol{x}_{c},t\right)\right\rangle .\label{ansatz-1}
\end{equation} In principle, the wavefunction in slowly perturbed crystal is a continuous superposition of Bloch states that has certain extension in momentum space and consequently also has a width in real space, thus pictured as a wavepacket. Since the spatial variation of the potential has a characteristic length much larger than the lattice constant, the wavepacket feels the periodicity of the potential and its extension in momentum space is sharply centred at some momentum $\boldsymbol{q}_{c}$. We then approximate the wavefunction by Eq.~(\ref{ansatz-1}) and understand it as a wavepacket parameterised by
the real-space centre $\boldsymbol{x}_{c}$
and the momentum-space centre $\hbar\boldsymbol{q}_{c}$, namely, $\left\vert \Phi\left(t\right)\right\rangle =\left\vert \Phi\left(\boldsymbol{q}_{c},\boldsymbol{x}_{c},t\right)\right\rangle$. Note here that in Eq.~(\ref{ansatz-1}), the band index $n$
is summed over all bands of $H_{c}\left(\boldsymbol{x}_{c},t\right)$
for the proper inclusion of non-adiabatic effects.

As the wavepacket can move throughout the space, its position $\boldsymbol{x}_{c}$
and momentum $\hbar\boldsymbol{q}_{c}$ can change over time. The
changing rate $\dot{\boldsymbol{x}}_{c}$ is naturally the velocity
and $\hbar\dot{\boldsymbol{q}}_{c}$ is interpreted as the force associated
with the rate of change of the crystal momentum. Using the Heisenberg
EOM, the velocity operator is found to be
\begin{equation}
\hat{\boldsymbol{V}}\left(\boldsymbol{x}_{c},t\right)=
-\frac{i}{\hbar}\left[\hat{\boldsymbol{X}},H_{c}\left(\boldsymbol{x}_{c},t\right)\right],\label{velo-opt-1}
\end{equation} The
expectation value of the velocity operator should be the wavepacket's
spatial velocity,
\begin{equation}
\dot{\boldsymbol{x}}_{c}=\left\langle \Phi\left(t\right)\right\vert \hat{\boldsymbol{V}}\left(\boldsymbol{x}_{c},t\right)\left\vert \Phi\left(t\right)\right\rangle ,\label{velo-exp-1}
\end{equation}
 on a self-consistent ground.

%This description for the velocity's expectation
%value is still meaningful when $H_{c}$ does not depends on $\boldsymbol{x}_{c}$,
%as the usual case of a spatially homogenous system.

In the Ehrenfest theorem, the so-called force operator is defined
by the minus of the potential gradient. Similarly, here we define a
force operator by,
\begin{equation}
\hat{\boldsymbol{F}}\left(\boldsymbol{x}_{c},t\right)=-\frac{\partial H_{c}\left(\boldsymbol{x}_{c},t\right)}{\partial\boldsymbol{x}_{c}}.\label{force-opt-1}
\end{equation}
 The expectation value of the Ehrenfest's force operator gives the
time derivative of the expectation value of the bare momentum operator.
In parallel to this, we replace the bare momentum
by the crystal momentum $\hbar\boldsymbol{q}_{c}$ in our case, and
its time derivative is then assumed to be the expectation value of
the corresponding force operator, namely,
\begin{equation}
\hbar\dot{\boldsymbol{q}}_{c}=\left\langle \Phi\left(t\right)\right\vert \hat{\boldsymbol{F}}\left(\boldsymbol{x}_{c},t\right)\left\vert \Phi\left(t\right)\right\rangle .\label{force-exp-1}
\end{equation}

Substituting Eq. (\ref{ansatz-1}) into Eqs. (\ref{velo-exp-1}) and
(\ref{force-exp-1}), we are led to
\begin{equation}
\dot{\boldsymbol{x}}_{c}=\left\langle \chi\left(t\right)\right\vert \frac{\partial\mathcal{H}^{c}\left(\boldsymbol{x}_{c},\boldsymbol{q}_{c},t\right)}{\partial\hbar\boldsymbol{q}_{c}}\left\vert \chi\left(t\right)\right\rangle ,\label{velo-exp-2}
\end{equation}

\begin{equation}
\hbar\dot{\boldsymbol{q}}_{c}=-\left\langle \chi\left(t\right)\right\vert \frac{\partial\mathcal{H}^{c}\left(\boldsymbol{x}_{c},\boldsymbol{q}_{c},t\right)}{\partial\boldsymbol{x}_{c}}\left\vert \chi\left(t\right)\right\rangle ,\label{force-exp-2}
\end{equation}
 where $\mathcal{H}^{c}\left(\boldsymbol{x}_{c},\boldsymbol{q}_{c},t\right)
 =e^{-i\boldsymbol{q}_{c}\cdot\hat{\boldsymbol{X}}}H_{c}\left(\boldsymbol{x}_{c},t\right)
 e^{i\boldsymbol{q}_{c}\cdot\hat{\boldsymbol{X}}}$. Here
 $\left\vert \chi\left(t\right)\right\rangle =\sum_{n}\eta_{n}\left(t\right)\left\vert u_{n,\boldsymbol{q}_{c}}\left(\boldsymbol{x}_{c},t\right)\right\rangle $
with $\left\vert u_{n,\boldsymbol{q}_{c}}\left(\boldsymbol{x}_{c},t\right)\right\rangle =e^{-i\boldsymbol{q}_{c}\cdot\hat{\boldsymbol{X}}}\left\vert \psi_{n,\boldsymbol{q}_{c}}\left(\boldsymbol{x}_{c},t\right)\right\rangle $
and it is subject to
\begin{equation}
\mathcal{H}^{c}\left(\boldsymbol{x}_{c},\boldsymbol{q}_{c},t\right)\left\vert \chi\left(t\right)\right\rangle =i\hbar\left\vert \dot{\chi}\left(t\right)\right\rangle .\label{locSchEq-2}
\end{equation}
The set of SC-EOM, Eqs.~(\ref{velo-exp-2}) and
(\ref{force-exp-2}), for the wavepacket's centre-of-mass $\left(\boldsymbol{x}_{c},\hbar\boldsymbol{q}_{c}\right)$ are
coupled to the quantum Schr\"{o}dinger
equation, Eq. (\ref{locSchEq-2}), which describes superposition among bands by the state vector, $\left\vert \chi\right\rangle$, quantified by the band amplitudes $\eta_{n}$'s.
The physical state of the wavepacket is thus completely specified by its centre-of-mass $(\boldsymbol{x}_{c},\hbar\boldsymbol{k}_{c})$
plus $\left\vert \chi\right\rangle$, here  called the band state. We will show how Eqs.~(\ref{velo-exp-2}) and
(\ref{force-exp-2}) can be recast into the more familiar gauge-covariant form Eqs.~(\ref{velo-nq-1}) and (\ref{force-nq-1}) in Sec.~\ref{wavpck-dys-1-2}.

%The latter is governed by the total time-dependence of $\mathcal{H}^{c}\left(\boldsymbol{x}_{c},\boldsymbol{q}_{c},t\right)$
%which is in turn dynamically parameterised by the centre-of-mass $\left(\boldsymbol{x}_{c},\hbar\boldsymbol{q}_{c}\right)$.

%If $H_{c}$ does not depend on $\boldsymbol{x}_{c}$, then the crystal
%momentum will not change. This anticipation is fully expressed by
%Eqs. (\ref{force-opt-1}) and (\ref{force-exp-1}) for the force operator
%itself vanishes in the absence of the spatial variation as the cause
%of force.

%\subsubsection{normal and anomalous velocities}
%\label{wavpck-dys-1-1}

The wavepacket's velocity, Eq.~(\ref{velo-exp-2}), can be decomposed in the band basis as
\begin{subequations}
\label{velo-sep-1}
\begin{equation}
\dot{\boldsymbol{x}}_{c}=\boldsymbol{v}_{b}+\boldsymbol{v}_{h},\label{velo-sep-1-tot}
\end{equation}
 where
\begin{equation}
\boldsymbol{v}_{b}\equiv\sum_{n}\eta_{n}^{*}\left\langle u_{n}\right\vert \frac{\partial\mathcal{H}^{c}}{\partial\hbar\boldsymbol{q}_{c}}\left\vert u_{n}\right\rangle \eta_{n}=\sum_{n}\left\vert \eta_{n}\right\vert ^{2}\frac{\partial\varepsilon_{n}}{\partial\hbar\boldsymbol{q}_{c}},\label{velo-sep-1-diag}
\end{equation}
 is the normal velocity associated with the dispersion of each band
and
\begin{align}
\boldsymbol{v}_{h} & \equiv\sum_{n}\sum_{m\ne n}\eta_{n}^{*}\left\langle u_{n}\right\vert \frac{\partial\mathcal{H}^{c}}{\partial\hbar\boldsymbol{q}_{c}}\left\vert u_{m}\right\rangle \eta_{m},
\label{velo-sep-1-offdiag}
\end{align}
 \end{subequations} is the anomalous velocity which is eventually
expressed in terms of Berry curvatures when we group bands into manifolds
(see later discussion). Here we have abbreviated $\left\vert u_{n}\right\rangle $
for $\left\vert u_{n,\boldsymbol{q}_{c}}\left(\boldsymbol{x}_{c},t\right)\right\rangle $ and $\varepsilon_{n}$ for $\varepsilon_{n,\boldsymbol{q}_{c}}\left(\boldsymbol{x}_{c},t\right)$ respectively.

Assuming initially the electron only occupies one particular band
$n_{0}$, namely, $\eta_{n}\left(t_{0}\right)=\delta_{n,n_{0}}$,
the system described by Eq. (\ref{locSchEq-1}) (and consequently
by Eqs. (\ref{velo-exp-2}), (\ref{force-exp-2}) and (\ref{locSchEq-2}))
under no external field will remain in just occupying that band $n_{0}$
rendering $\eta_{n}\left(t\right)=0$ whenever $n\ne n_{0}$. In such
a trivial situation, the anomalous velocity $\boldsymbol{v}_{h}$
would be zero. The anomalous motion, in contrast to the normal one,
is thus a result of inter-band transition induced by the external
field. In addition, for transitions between a pair of degenerate bands
$\left(n,m\right)$ with $n\ne m$ and $\varepsilon_{n}=\varepsilon_{m}\equiv\varepsilon^{d}$, we have in general
\begin{align}
\label{deg-vh}
\left\langle u_{n}\right\vert\frac{\partial\mathcal{H}^{c}}{\partial\hbar\boldsymbol{q}_{c}}\left\vert u_{m}\right\rangle=\varepsilon^{d}\frac{\partial}{\partial\boldsymbol{q}_{c}}\left\langle u_{n}\right\vert\left.u_{m}\right\rangle =\varepsilon^{d}\frac{\partial\delta_{n,m}}{\partial\boldsymbol{q}_{c}}=0.
\end{align} Only transitions between non-degenerate bands in Eq.~(\ref{velo-sep-1-offdiag})
contribute to the anomalous velocity.

\subsubsection{Gauge invariance of expectation values of observables and gauge
covariance form of EOM}
\label{wavpck-dys-1-2}

In principle, the motion of a wavepacket in multiple bands
can be studied using Eqs. (\ref{velo-exp-2}), (\ref{force-exp-2})
and (\ref{locSchEq-2}). To verify their validity, we will connect
them with the more familar form of SC-EOM.\cite{Culcer05085110,Chang967010,Sundaram9914915}
in which the Berry curvatures explicitly appear. First we define the Berry connections,
\begin{equation}
\left[\mathcal{\mathcal{R}}_{\lambda_{\alpha}}\right]_{n,m}=\left\langle u_{n}\left\vert i\frac{\partial u_{m}}{\partial\lambda_{\alpha}}\right.\right\rangle ,\label{BerryCon-def-comp}
\end{equation} in the phase space $\boldsymbol{\lambda}=\left(\boldsymbol{x}_{c},\hbar\boldsymbol{q}_{c}\right)$,
 where $\lambda_{\alpha}$ stands for the $\alpha$-th component of
$\boldsymbol{\lambda}$ . Using the decomposition Eq. (\ref{velo-sep-1})
with the observation, $\sum_{n}\sum_{m\ne n}\eta_{n}^{*}\left\langle u_{n}\right\vert \frac{\partial\mathcal{H}^{c}}{\partial\hbar\boldsymbol{q}_{c}}\left\vert u_{m}\right\rangle \eta_{m}=
-i\sum_{n}\sum_{m\ne n}\eta_{n}^{*}\left[\mathcal{R}_{\hbar\boldsymbol{q}_{c}},\mathcal{H}^{c}\right]_{n,m}\eta_{m}$,
 and applying similar decomposition to Eq. (\ref{force-exp-2}), then Eqs.
(\ref{velo-exp-2}), (\ref{force-exp-2}) and (\ref{locSchEq-2})
are rewritten into
\begin{equation}
\dot{\boldsymbol{x}}_{c}=\left\langle \left[\mathcal{D}_{\hbar\boldsymbol{q}_{c}},\mathcal{H}^{c}\right]\right\rangle ,\label{velo-nq-1}
\end{equation}

\begin{equation}
\hbar\dot{\boldsymbol{q}}_{c}=-\left\langle \left[\mathcal{D}_{\boldsymbol{x}_{c}},\mathcal{H}^{c}\right]\right\rangle ,\label{force-nq-1}
\end{equation}
 and
\begin{equation}
i\hbar\frac{D}{Dt}\boldsymbol{\eta}\left(t\right)=\mathcal{H}^{c}\left(t\right)\boldsymbol{\eta}\left(t\right),\label{locSchEq-3}
\end{equation}
 respectively where
\begin{equation}
\left[\mathcal{D}_{\lambda_{\alpha}}\right]_{n,m}=\delta_{n,m}\frac{\partial}{\partial\lambda_{\alpha}}-i\left[\mathcal{\mathcal{R}}_{\lambda_{\alpha}}\right]_{n,m},\label{covD-1}
\end{equation}

\begin{equation}
\left[\frac{D}{Dt}\right]_{n,m}=\delta_{n,m}\frac{d}{dt}-i\sum_{\alpha}\left[\mathcal{\mathcal{R}}_{\lambda_{\alpha}}\right]_{n,m}\dot{\lambda}_{\alpha},\label{covD-t}
\end{equation}
Here
$\left\langle O\right\rangle =\sum_{n,m}\eta_{n}^{*}O_{n,m}\eta_{m}$
for any matrix quantity $O$ and $\boldsymbol{\eta}$ the
column vector with entries $\eta_{n}$'s. The
expression Eq.~(\ref{velo-nq-1}) bares the interpretation of
gauge covariant group velocity,\cite{Culcer05085110} due to the appearance
of the so-called covariant derivative, Eq.~(\ref{covD-1}). From the above derivations, we see that this gauge covariant group velocity Eq.~(\ref{velo-nq-1}) is exactly
the expectation value of the quantum velocity operator under the dynamics
of the wavepacket's local Hamiltonian, Eq.~(\ref{velo-exp-2}). The
equality between Eq. (\ref{velo-nq-1}) and Eq.~(\ref{velo-exp-2})
and the equality between Eq. (\ref{force-nq-1}) and Eq.~(\ref{force-exp-2})
simply manifest the gauge invariance of the expectation values of
physical observables described by gauge covariant EOM.

The more familiar form of Eq.~(\ref{force-nq-1})
or Eq. (\ref{force-exp-2}) explicitly contains a term proportional
to the external electric field $\boldsymbol{E}\left(\boldsymbol{x}_{c},t\right)=-\partial\boldsymbol{A}/\partial t-\partial\delta\phi/\partial\boldsymbol{x}_{c}$, where $\delta{\phi}$ is the externally applied scalar potential that smoothly changes in space. This
is easily arrived by the translational invariance that relates the
Bloch state for momentum $\hbar\boldsymbol{q}_{c}$ via the vector
potential $\boldsymbol{A}$ to $\hbar\boldsymbol{k}_{c}$
by $\hbar\boldsymbol{k}_{c}=\hbar\boldsymbol{q}_{c}-\left(-e\right)\boldsymbol{A}$ in which $\hbar\boldsymbol{k}_{c}$ is called mechanical crystal momentum.~\cite{Sundaram9914915} We further
replace $\partial/\partial\boldsymbol{q}_{c}$ appearing in $\dot{\boldsymbol{x}}_{c}$
 by $\partial/\partial\boldsymbol{k}_{c}$ according to
the chain rule, resulting in\cite{note-1}
\begin{equation}
\dot{\boldsymbol{x}}_{c}=\left\langle \left[\mathcal{D}_{\hbar\boldsymbol{k}_{c}},\mathcal{H}^{c}\right]\right\rangle =\left\langle \chi\right\vert \frac{\partial\mathcal{H}^{c}}{\partial\hbar\boldsymbol{k}_{c}}\left\vert \chi\right\rangle ,\label{velo-nq-2}
\end{equation}
and
\begin{equation}
\hbar\dot{\boldsymbol{k}}_{c}=(-e)\boldsymbol{E}-\left\langle \left[\mathcal{D}_{\boldsymbol{x}_{c}},\mathcal{H}^{c}\right]\right\rangle =(-e)\boldsymbol{E}
+\boldsymbol{F}_{c},\label{force-nq-2}
\end{equation}
where
\begin{equation}
\boldsymbol{F}_{c}=-\left\langle \chi\right\vert \frac{\partial\mathcal{H}^{c}}{\partial\boldsymbol{x}_{c}}\left\vert \chi\right\rangle
\label{force-nq-2g}
\end{equation} and
$\mathcal{H}^{c}$ in Eqs. (\ref{velo-nq-2}) and (\ref{force-nq-2})
is
\begin{equation}
\mathcal{H}^{c}\left(\boldsymbol{x}_{c},\boldsymbol{k}_{c}\right)=e^{-i\boldsymbol{k}_{c}\cdot\hat{\boldsymbol{X}}}\left.H_{c}\left(\boldsymbol{x}_{c},t\right)\right\vert _{\boldsymbol{A}=0,\delta\phi=0}e^{i\boldsymbol{k}_{c}\cdot\hat{\boldsymbol{X}}},\label{Hc0-1}
\end{equation} and
does not explicitly depends on $t$ since the explicit $t-$dependence
in $H_{c}\left(\boldsymbol{x}_{c},t\right)$ solely comes from the
electric-field-generating vector potential. Note that the trivial case of spatial variations in which all bands' energies change by the same amount in space has already been taken care of by $\delta\phi$ so $\boldsymbol{E}$ can be dependent on $\boldsymbol{x}_{c}$.

In the ideal case of no spatial perturbation as what occurs in a perfect periodic lattice without electromagnetic field, the local Schr\"{o}dinger equation becomes globally valid and $\partial H_{c}\left(\boldsymbol{x}_{c},t\right)/\partial\boldsymbol{x}_{c}=0$. The description
by Eq.~(\ref{locSchEq-1}) becomes a standard quantum mechanics problem
for a particle moving in a periodic potential. Nevertheless, the velocity observable is still well-defined by the operator Eq.~(\ref{velo-opt-1}). The system that starts
with a momentum $\hbar\boldsymbol{q}_{c}$ as a delocalised Bloch
state will remain delocalised with the velocity given by the right-hand sides of Eq.~(\ref{velo-exp-2})
and equally by Eq.~(\ref{velo-nq-1}) but without the semiclassical
notion as a localised wavepacket.

\subsection{Dynamics of a wavepacket within the active manifold}

\label{wavpck-dys-2}

Eqs. (\ref{velo-nq-1}) and (\ref{force-nq-1}) already resemble
the non-Abelian SC-EOM used in Refs.~\cite{Culcer05085110,Xiao101959} For pedagogical reasons, below we continue to discuss how the Berry curvatures emerge from Eqs.~(\ref{velo-nq-1}) and (\ref{force-nq-1}) or equivalently from Eqs.~(\ref{velo-exp-2}) and (\ref{force-exp-2}).

We denote the active manifold by $a$ and the rest by $r$. With these labels of the bands, we can compactly rewrite
Eqs. (\ref{velo-nq-1}) and (\ref{force-nq-1}) into
\begin{align}
\dot{\lambda}_{\alpha} & =\text{sign}\left(\lambda_{\alpha}\right)\left\{ \left\langle \left[\mathcal{D}_{\hat{\lambda}_{\alpha}},\mathcal{H}_{a}^{c}\right]\right\rangle_{a}+\left\langle \left[\mathcal{D}_{\hat{\lambda}_{\alpha}},\mathcal{H}_{r}^{c}\right]\right\rangle _{r}\right.\nonumber \\
 & \left.+\left(\sum_{n\in a}\sum_{l\in r}\eta_{n}^{*}\left\langle u_{n}\right\vert \frac{\partial\mathcal{H}^{c}}{\partial\hat{\lambda}_{\alpha}}\left\vert u_{l}\right\rangle \eta_{l}+\text{c.c.}\right)\right\} ,\label{phsvelo-sep-1}
\end{align}
 where $\lambda_{\alpha}$ is the $\alpha$-th component ( in terms
of spatial direction) of $\boldsymbol{x}_{c}$ or $\hbar\boldsymbol{q}_{c}$, $\hat{\lambda}_{\alpha}$ stands for the conjugate, namely, $\hat{\boldsymbol{x}}_{c}=\hbar\boldsymbol{q}_{c}$
and $\hbar\hat{\boldsymbol{q}}_{c}=\boldsymbol{x}_{c}$ , and the
sign function values as $\text{sign}\left(\boldsymbol{x}_{c}\right)=1$
and $\text{sign}\left(\hbar\boldsymbol{q}_{c}\right)=-1$. We use
the notation $\left\langle O\right\rangle _{a/r}$ for averages only
over the bands in the manifold $a/r$ and $\mathcal{H}_{a/r}^{c}$
for the block of $\mathcal{H}^{c}$ in the space of
$a/r$.

The reliability of the following approximations underlie the meaning of grouping the bands into $a$ and $r$.~\cite{Thouless82405,Boehm03book}
(i) The occupation of $r$ bands is negligible,
corresponding to ignore the term $\left\langle \left[\mathcal{D}_{\hat{\lambda}_{\alpha}},\mathcal{H}_{r}^{c}\right]\right\rangle _{r}$
of Eq. (\ref{phsvelo-sep-1}). (ii) The coherence between $a$ and $r$,
described by the second line of Eq. (\ref{phsvelo-sep-1}) shall be
kept up to the lowest order of the external field. So leading-order effects of $r$ on $a$ survive. Note that the effect of the external field on transitions within the active manifold is kept to all orders.

After these approximations, Eq.~(\ref{phsvelo-sep-1})
is then turned into Eq.~(\ref{phsvelo-sep-2}) with the Berry curvature defined in Eq.~(\ref{nABCV-1}) (see the derivation in the followings).

\subsubsection{Emergence of Berry curvatures}
\label{wavpck-dys-2-2}

To see how Berry curvatures arise in the second line of Eq.~(\ref{phsvelo-sep-1}), we have to investigate the coherent dynamics of $\boldsymbol{\eta}\left(t\right)$, under an electric field which is relatively small with respect to inter-manifold energy separation. Note that the coherence between bands $n$ and $m$ is characterised by the relative phase, $\text{arg}\left(\eta_{n}^{*}\eta_{m}/\left\vert\eta_{n}^{*}\eta_{m}\right\vert\right)$. Naively solving Eq.~(\ref{locSchEq-3}) leads to gauge-dependent relative phases.~\cite{note-2-1} This complication is avoided by re-inspecting Eq.~(\ref{locSchEq-2}) with $\left\vert \chi\right\rangle=\sum_{n}\bar{\eta}_{n}\left\vert\bar{u}_{n}\right\rangle$, where
\begin{subequations}
\label{ginvinsbasis}
\begin{align}
\left\vert\bar{u}_{n}\right\rangle
=e^{i\gamma_{n}}\left\vert{u}_{n}\right\rangle,
\label{ginv-bvec}
\end{align} and
\begin{align}
\bar{\eta}_{n}=e^{-i\gamma_{n}}\eta_{n},
\label{ginv-amp}
\end{align}
\end{subequations}in which $\gamma_{n}$ is the Berry phase defined by
\begin{align}
\gamma_{n}=\int^{\left(\boldsymbol{x}_{c},\boldsymbol{k}_{c}\right)}_{\left(\boldsymbol{x}_{0},\boldsymbol{k}_{0}\right)}\left\{ \text{d}\boldsymbol{x}^{\prime}\cdot\left[\mathcal{R}_{\boldsymbol{x}^{\prime}}\right]_{n,n}+\text{d}\boldsymbol{k}^{\prime}\cdot
\left[\mathcal{R}_{\boldsymbol{k}^{\prime}}\right]_{n,n}\right\}.
\end{align} here $\left(\boldsymbol{x}_{0},\boldsymbol{k}_{0}\right)$ denotes the initial value of the wavepacket's centre-of-mass. This results in
\begin{align}
\bar{\mathcal{H}}\bar{\boldsymbol{\eta}}
=i\hbar\dot{\bar{\boldsymbol{\eta}}},\label{effSCE-1}
\end{align}  where $\bar{\boldsymbol{\eta}}$ denotes the vector with entries $\bar{\eta}_{n}$'s and the matrix $\bar{\mathcal{H}}$ has elements,
\begin{equation}
\bar{\mathcal{H}}_{n,m}=\delta_{n,m}\varepsilon_{n}+\left(1-\delta_{n,m}\right)V_{n,m},\label{mvfrm-H-0}
\end{equation}
with
\begin{equation}
V_{n,m}=-\hbar\left\{ \left[\bar{\mathcal{R}}_{\boldsymbol{x}_{c}}\right]_{n,m}\cdot\dot{\boldsymbol{x}}_{c}
+\left[\bar{\mathcal{R}}_{\boldsymbol{k}_{c}}\right]_{n,m}\cdot\dot{\boldsymbol{k}}_{c}\right\}
\label{hyb-coup}
\end{equation} in which $\left[\bar{\mathcal{R}}_{\boldsymbol{x}_{c}}\right]_{n,m}=e^{-i\gamma_{n}}\left[\mathcal{R}_{\boldsymbol{x}_{c}}\right]_{n,m}e^{i\gamma_{m}}$ with a similar definition applied to $\left[\bar{\mathcal{R}}_{\boldsymbol{k}_{c}}\right]_{n,m}$. Here $\bar{\mathcal{H}}$ is effectively a Hamiltonian in the moving frame of the carrier, whose motion in the phase space due to the electric field and the spatial variation of the band structures induces inter-band coherent hybridisation, mediated by the matrix elements with $n\ne m$ in Eq.~(\ref{mvfrm-H-0}).~\cite{Tu19}

%%%%%%%%%%%%%%%%%%%%%%%%%%%%%%%%%%%%%%%%%%%%%%%%%%%%%%%%%%%%%%%%%%%%%%%%%%%%%%%-s

We denote the projection of $\bar{\boldsymbol{\eta}}$ on the $a$-bands by $\bar{\boldsymbol{\eta}}_{a}$ and those on $r$ by $\bar{\boldsymbol{\eta}}_{r}$, namely,
$\bar{\boldsymbol{\eta}}=\left(\begin{array}{c}
\bar{\boldsymbol{\eta}}_{a}\\
\bar{\boldsymbol{\eta}}_{r}
\end{array}\right)$, and Eq.~(\ref{effSCE-1}) becomes
\begin{align}
\left(\begin{array}{cc}
\bar{\mathcal{H}}_{a} & V_{a,r}\\
V_{r,a} & \bar{\mathcal{H}}_{r}
\end{array}\right)\left(\begin{array}{c}
\bar{\boldsymbol{\eta}}_{a}\\
\bar{\boldsymbol{\eta}}_{r}
\end{array}\right)=i\hbar\left(\begin{array}{c}
\dot{\bar{\boldsymbol{\eta}}}_{a}\\
\dot{\bar{\boldsymbol{\eta}}}_{r}
\end{array}\right),
\end{align}
 where $\bar{\mathcal{H}}_{a}$ ($\bar{\mathcal{H}}_{r}$) is the $a$($r$)-block of the effective Hamiltonian $\bar{\mathcal{H}}$ in Eq.~(\ref{mvfrm-H-0})
and $V_{a,r}=-\hbar\bar{\mathcal{K}}^{a,r}$ with $\left[\bar{\mathcal{K}}^{a,r}\right]_{n,l}\equiv \left\{ \left[\bar{\mathcal{R}}_{\boldsymbol{x}_{c}}\right]_{n,l}\cdot\dot{\boldsymbol{x}}_{c}
+\left[\bar{\mathcal{R}}_{\boldsymbol{k}_{c}}\right]_{n,l}\cdot\dot{\boldsymbol{k}}_{c}\right\} $
for $n\in a$ and $l\in r$ with $V_{r,a}=-\hbar\bar{\mathcal{K}}^{r,a}$
the hermitian conjugate of $V_{a,r}$. It
is convenient to work in the rotating frame by defining
\begin{equation}
\tilde{\boldsymbol{\eta}}\left(t\right)=U_{0}^{\dagger}\left(t\right)\bar{\boldsymbol{\eta}}\left(t\right),
\end{equation}
 where
\begin{equation}
U_{0}\left(t\right)=\left(\begin{array}{cc}
U_{a}\left(t\right) & 0\\
0 & U_{r}\left(t\right)
\end{array}\right),
\end{equation}
 with $U_{a/r}\left(t\right)=\hat{T}\exp\left\{ -\frac{i}{\hbar}\int_{t_{0}}^{t}\text{d}\tau\bar{\mathcal{H}}_{a/r}\left(\tau\right)\right\} $
in which $\hat{T}$ is the time ordering operator and $t_{0}$ is
the initial time. Since the strength of the coupling $V_{a,r}$
is inversely proportional to the large energy separation between $a$
and $r$, it is treated as a perturbation. Up to the leading order,
we have
\begin{equation}
\dot{\tilde{\boldsymbol{\eta}}}\left(t\right)\approx i\bar{\mathcal{K}}_{\text{mix}}\tilde{\boldsymbol{\eta}}\left(t_{0}\right),\label{adiapert-1}
\end{equation}
 where
\begin{equation}
\bar{\mathcal{K}}_{\text{mix}}=\left(\begin{array}{cc}
0 & U_{a}^{\dagger}\bar{\mathcal{K}}^{a,r}U_{r}\\
U_{r}^{\dagger}\bar{\mathcal{K}}^{r,a}U_{a} & 0
\end{array}\right).
\end{equation}
 We assume initially there is no occupation on $r$-bands,
namely, $\bar{\boldsymbol{\eta}}_{r}\left(t_{0}\right)=0$. This initial
condition together with the approximation Eq. (\ref{adiapert-1})
results in
\begin{equation}
i\hbar\dot{\bar{\boldsymbol{\eta}}}_{a}=\bar{\mathcal{H}}_{a}\bar{\boldsymbol{\eta}}_{a},\label{eta-d-lgg-1}
\end{equation} which describes the coherent dynamics of the band amplitudes within the active manifold, also called the \textit{band pseudospin}. We are interested only in the inter-band transitions among the bands
in $a$ and we neglect the inter-band transitions among the bands in $r$
leading to
\begin{align}
\dot{\tilde{\eta}}_{l\in r} & =i\sum_{m\in a}e^{\left(i/\hbar\right)\int_{t_{0}}^{t}\text{d}s\left(\varepsilon_{l}\left(s\right)-\varepsilon_{m}\left(s\right)\right)}
\left[\bar{\mathcal{K}}^{r,a}\right]_{lm}\tilde{\eta}_{m}.\label{eta-s-lgg-1}
\end{align} Due to the large gap, the exponential factor oscillates very fast
while the adiabatic approximation gives that the inter-manifold coupling $\left[\bar{\mathcal{K}}^{r,a}\right]_{lm}$
changes slowly. Following the adiabatic approximations in Ref.
{[}\onlinecite{Xiao101959}{]}, we can perform integration by part
on Eq. (\ref{eta-s-lgg-1}), yielding
\begin{equation}
\bar{\eta}_{l}=\sum_{m\in a}\frac{\hbar\left[\bar{\mathcal{K}}^{r,a}\right]_{lm}}{\left(\varepsilon_{l}-\varepsilon_{m}\right)}
\bar{\eta}_{m}.\label{adia-mb-rl}
\end{equation}
Defining the dimensionless factors,
\begin{align}
\epsilon=\left\vert \frac{\hbar\left[\bar{\mathcal{K}}^{r,a}\right]_{lm}}{\left(\varepsilon_{l}-\varepsilon_{m}\right)}\right\vert,
\label{epsilon}
\end{align}
we see the average on the $r$-bands, $\left\langle \left[\mathcal{D}_{\hat{\lambda}_{\alpha}},\mathcal{H}_{r}^{c}\right]\right\rangle _{r}$,
is two orders of $\epsilon$ smaller than $\left\langle \left[\mathcal{D}_{\hat{\lambda}_{\alpha}},\mathcal{H}_{a}^{c}\right]\right\rangle _{a}$
and the occupations on the $r$-bands are thus neglected.
Substituting Eq.~(\ref{adia-mb-rl}) into Eq.~(\ref{phsvelo-sep-1}) with the aid of Eq.~(\ref{ginvinsbasis}),
the contribution from transitions between $a$ and $r$ to the velocity $\dot{\boldsymbol{x}}_{c}$ reads,
\begin{widetext}
\begin{align}
\left(\sum_{n\in a}\sum_{l\in r}\eta_{n}^{*}\left\langle u_{n}\right\vert \frac{\partial\mathcal{H}^{c}}{\partial\hat{\lambda}_{\alpha}}\left\vert u_{l}\right\rangle \eta_{l}+\text{c.c.}\right) & =-i\hbar\sum_{\beta}\sum_{n\in a}\sum_{l\in r}\eta_{n}^{*}\sum_{m\in a}\frac{\left(\varepsilon_{l}-\varepsilon_{n}\right)}{\left(\varepsilon_{l}-\varepsilon_{m}\right)}\left\langle \left.\frac{\partial u_{n}}{\partial\hat{\lambda}_{\alpha}}\right\vert u_{l}\right\rangle \left\langle u_{l}\left\vert \frac{\partial u_{m}}{\partial\lambda_{\beta}}\right.\right\rangle \dot{\lambda}_{\beta}\eta_{m}+\text{c.c.}\label{off-trasit-0}
\end{align}
 \end{widetext} where $\hat{\lambda}_{\alpha}$ is the $\alpha$th component of $\hbar\boldsymbol{k}_{c}$. The same expression applies to the contribution from the $a$-$r$ transition to the force $\hbar\dot{\boldsymbol{k}}_{c}$ with $\hat{\lambda}_{\alpha}$ replaced by the $\alpha$th component of $\boldsymbol{x}_{c}$. The remoteness of bands in $r$ from $a$ assures
the validity of the approximation,
\begin{equation}
\frac{\left(\varepsilon_{l}-\varepsilon_{n}\right)}{\left(\varepsilon_{l}-\varepsilon_{m}\right)}\approx1,\label{r-d-approx1}
\end{equation} without requiring exact degeneracy among the bands in $a$. Furthermore, we observe that
\begin{equation}
i\left\langle \frac{\partial u_{n}}{\partial\hat{\lambda}_{\alpha}}\right\vert \left[\sum_{l\in r}\left\vert u_{l}\right\rangle \left\langle u_{l}\right\vert \right]\left\vert\frac{\partial u_{m}}{\partial\lambda_{\beta}}\right\rangle +\text{c.c.}=\mathcal{F}_{nm}^{\hat{\lambda}_{\alpha}\lambda_{\beta}},\label{nABCV-0}
\end{equation}
is the non-Abelian Berry curvature matrix that can be rewritten in a more familiar form,
\begin{equation}
\mathcal{F}_{nm}^{\hat{\lambda}_{\alpha}\lambda_{\beta}}=\left\{ \frac{\partial\left[\mathcal{\mathcal{R}}_{\lambda_{\beta}}\right]_{nm}}{\partial\hat{\lambda}_{\alpha}}
-\frac{\partial\left[\mathcal{\mathcal{R}}_{\hat{\lambda}_{\alpha}}\right]_{nm}}{\partial\lambda_{\beta}}
-i\left[\mathcal{\mathcal{R}}_{\hat{\lambda}_{\alpha}},\mathcal{\mathcal{R}}_{\lambda_{\beta}}\right]_{nm}\right\}.\label{nABCV-1}
\end{equation} with the help of Eq.~(\ref{BerryCon-def-comp}). Using the approximation Eq.~(\ref{r-d-approx1}) in Eq.~(\ref{off-trasit-0})
for Eq.~(\ref{phsvelo-sep-1}) with the identification of Eq.~(\ref{nABCV-0}) with Eq.~(\ref{nABCV-1}),
we finally see that Eqs.~(\ref{velo-exp-2}) and (\ref{force-exp-2}) or equivalently Eqs.~(\ref{velo-nq-1}) and (\ref{force-nq-1}) (summarised as Eq.~(\ref{phsvelo-sep-1})) reduce to the non-Abelian SC-EOM in Ref. {[}\onlinecite{Culcer05085110}{]}, namely
\begin{subequations}
\label{phsvelo-sep-2}
\begin{align}
\dot{x}_{\alpha}\!=\!\left\langle\left[\mathcal{D}_{\hbar{k}_{\alpha}},\mathcal{H}_{a}^{c}\right]\right\rangle_{a}
\!-\!\!\sum_{\beta}\!\left(\left\langle\mathcal{F}^{{k}_{\alpha}x_{\beta}}\right\rangle_{a}\!\dot{x}_{\beta}
\!+\!\left\langle\mathcal{F}^{{k}_{\alpha}k_{\beta}}\right\rangle_{a}\!\dot{k}_{\beta}\right),\label{phsvelo-sep-2-x}
\end{align}
\begin{align}
\hbar\dot{k}_{\alpha}\!&=\!-eE_{\alpha}-\left\langle\left[\mathcal{D}_{{x}_{\alpha}},\mathcal{H}_{a}^{c}\right]\right\rangle_{a}
\nonumber\\&
+\hbar\sum_{\beta}\!\left(\left\langle\mathcal{F}^{{x}_{\alpha}x_{\beta}}\right\rangle_{a}\!\dot{x}_{\beta}
\!+\!\left\langle\mathcal{F}^{{x}_{\alpha}k_{\beta}}\right\rangle_{a}\!\dot{k}_{\beta}\right).\label{phsvelo-sep-2-k}
\end{align}
\end{subequations} We have omitted the subscript $c$ for the centre-of-mass variables, writing $\left(x_{c\alpha},k_{c\alpha}\right)$ simply by $\left(x_{\alpha},k_{\alpha}\right)$.

%And Using Previously Defined $\Left[\Mathcal{K}^{R,D}\Right]_{Lm}$ With The Notational Abbreviations Of $\Lambda_{\Alpha}$ For The $\Alpha$-Th Component Of $\Boldsymbol{X}_{C}$ And $\Hbar\Boldsymbol{K}_{C}$,

%%%%%%%%%%%%%%%%%%%%%%%%%%%%%%%%%%%%%%%%%%%%%%%%%%-e

We thus have deduced the SC-EOM with explicit
appearance of Berry curvatures, Eq.~(\ref{phsvelo-sep-2}), in a proper
limit from the full-band dynamics, namely Eqs. (\ref{velo-exp-2}) and (\ref{force-exp-2}),
or compactly as Eq. (\ref{phsvelo-sep-1}). The Abelian formulation\cite{Chang967010,Sundaram9914915}
is obtained from Eq. (\ref{phsvelo-sep-2}) by letting the manifold
$a$ contain only one band, resulting in
\begin{subequations}
\label{phsvelo-sep-2-AB}
\begin{equation}
\dot{x}_{\alpha}= \frac{\partial\varepsilon_{n_{}}}{\partial\hbar{k}_{\alpha}}-
\sum_{\beta}\left(\Omega_{n_{}}^{{k}_{\alpha}x_{\beta}}\dot{x}_{\beta}
+\Omega_{n_{}}^{{k}_{\alpha}k_{\beta}}\dot{k}_{\beta}\right)
,\label{phsvelo-sep-2-AB-x}
\end{equation}
\begin{equation}
\hbar\dot{k}_{\alpha}=-eE_{\alpha} -\frac{\partial\varepsilon_{n_{}}}{\partial{x}_{\alpha}}+\hbar
\sum_{\beta}\left(\Omega_{n_{}}^{{x}_{\alpha}x_{\beta}}\dot{x}_{\beta}
+\Omega_{n_{}}^{{x}_{\alpha}k_{\beta}}\dot{k}_{\beta}\right)
,\label{phsvelo-sep-2-AB-k}
\end{equation}
\end{subequations}
 where
\begin{equation}
\Omega_{n_{}}^{{\lambda}_{\alpha}\lambda_{\beta}}=
\frac{\partial\left[\mathcal{\mathcal{R}}_{\lambda_{\beta}}\right]_{n_{}n_{}}}{\partial{\lambda}_{\alpha}}-
\frac{\partial\left[\mathcal{\mathcal{R}}_{{\lambda}_{\alpha}}\right]_{n_{}n_{}}}{\partial\lambda_{\beta}}
\label{BCV-1}
\end{equation}
 is the Abelian Berry curvature for the band indexed by $n_{}$. The second terms in Eqs.~(\ref{phsvelo-sep-2}) and (\ref{phsvelo-sep-2-AB}) in terms of the Berry curvatures are of $\mathcal{O}(\epsilon)$, where $\epsilon$ is defined by Eq.~(\ref{epsilon}).

%Note that the usual meaning of the adiabatic limit refers to the case
%where the time-evolving state remains strictly in one band (or several
%degenerate bands) without considering any effects of transition
%to other bands. This is usually known as the zeroth order adiabatic
%approximation. The results of Eq. (\ref{phsvelo-sep-2}) are thus
%beyond the zeroth order adiabatic approximation by including the first-order
%non-adiabatic correction. To keep the terminology brief in the present
%context, we simply term Eqs. (\ref{phsvelo-sep-2})
%as the adiabatic limit throughout this article.

\subsubsection{Non-adiabaticity within the active manifold}
\label{wavpck-dys-2-3}

The SC-EOM, Eq.~(\ref{phsvelo-sep-2}), for the wavepacket dynamics within the active manifold already cover both the adiabatic (single band or several degenerate bands) and the non-adiabatic (several non-degenerate bands) cases, including effects from the momentum as well as the spatial textures. Degeneracy among active bands brings forth $U\left(N\right)$ symmetry, where $N$ is the number of bands within the active manifold and enables formal connection to non-Abelian gauge structure.~\cite{Wilczek842111} The second term of Eq.~(\ref{phsvelo-sep-2}) as the non-Abelian curvature has been a focus of previous discussions.\cite{Xiao101959,Culcer05085110,Shindou05399} To relate wavepacket dynamics to geometric effects arising from closed trajectories over the phase space with $U\left(N\right)$ symmetry, keeping degeneracy along the trajectories is thus presumed.~\cite{Wilczek842111} Note that Eq.~(\ref{phsvelo-sep-2}) makes no strict restrictions on energy spacing among the active bands over the phase space as long as they are far enough separated from $r$-bands for Eq.~(\ref{r-d-approx1}) to be a good approximation.

Here we want to address the first term of Eq.~(\ref{phsvelo-sep-2}) that contains non-adiabatic effects within the active manifold. Interestingly, from Eqs.~(\ref{velo-sep-1}) and (\ref{deg-vh}), degeneracy makes zero contributions to the anomalous velocity. Nonzero contribution to $\boldsymbol{v}_{h}$ only comes from inter-manifold transitions, as the second term of Eq.~(\ref{phsvelo-sep-2}), and crucially intra-manifold transitions among non-degenerate active bands, embedded in the first term of Eq.~(\ref{phsvelo-sep-2}). Recall Sec.~\ref{wavpck-dys-2-2} that the emergence of the second term of Eq.~(\ref{phsvelo-sep-2}) as non-Abelian Berry curvatures, is from a perturbation correction to the first order of a small parameter, $\epsilon$, Eq.~(\ref{epsilon}). Its contribution is thus on the order $\mathcal{O}(\epsilon)$. In contrast, the first term of Eq.~(\ref{phsvelo-sep-2}), that incorporates non-adiabatic effects, is on the order of $\epsilon^{0}$.
The consequence of this difference between adiabatic and non-adiabatic dynamics for multiple active bands on transport current will be further explored in Sec.~\ref{semiCcrnt-1-3}.

A number of physical realisations exists for manifesting the importance of the non-adiabatic dynamics. This includes materials with narrow band gaps whose sizes are comparable to the electric field, for example, those featuring band anti-crossings such as gapped Dirac cones~\cite{Sinova04126603,Zhang19206401,Li19121103,Li19eaaw5685,Liu2005733} and gap sign reversal in moir\'{e} patterns.~\cite{Hunt131427,Woods14451} Although the form of Eq.~(\ref{phsvelo-sep-2}) has already appeared in the literatures, the implications of the non-adiabatic contents of such EOM for a single wavepacket have not been fully explored in terms of transport current for an ensemble of wavepackets. Below, we continue to discuss the steady-state
transport within the semiclassical picture. We assume the electric field no longer changes with time in the steady-state limit.

%, where we do not want to have any restrictions imposed on
%transitions among bands,

\section{Semiclassical transport theories}

\label{semiCcrnt-1}

A semiclassical transport theory is featured by the capability of calculating the transport current within the phase-space framework for an ensemble of electron wavepackets forming an electron gas. The single-waveapacket basis for the description of the ensemble is the SC-EOM. Each manifold has its corresponding SC-EOM. Summing over contributions from the relevant manifolds (e.g. the lowest conduction band and the top valence band, in the case of a large gap semiconductor) then gives the transport current for the ensemble.

A well-established semiclassical transport theory is the standard kinetic theory,~\cite{Ashcroft76book} which relies on the availability of single-band manifolds with the SC-EOM given by Eq.~(\ref{phsvelo-sep-2-AB}). The expression for the current is the familiar form, Eq.~(\ref{Bltzeq-2}) (see details in Sec.~\ref{semiCcrnt-1-1-2}). This well-established transport theory based on single-band manifolds is referred here as single-band semiclassical transport theory (SSCT).  For being self-content, we first review in Sec.~\ref{semiCcrnt-1-1} the known results of SSCT and its correspondence to the pure classical picture of a phase-space fluid (PSF). The original picture of PSF is important in understanding the semiclassical transport current in SSCT.

When one relevant manifold consists of multiple bands, the corresponding SC-EOM is given by Eq.~(\ref{phsvelo-sep-2}), regardless it is degenerate or not. The transport theory for such multiple-band manifold is referred here as multiple-band semiclassical transport theory (MSCT). In Sec.~\ref{semiCcrnt-1-2}, we first show that the nature of having multiple bands, regardless of being degenerate or not, renders it difficult to directly extend from SSCT to MSCT. Nevertheless, under restricted conditions (see Sec.~\ref{semiCcrnt-1-2-3} for details), straightforward generalisation from SSCT to MSCT results in a familiar expression of current, Eq.~(\ref{anom-crnt-Ad-nAB-0}), in terms of non-Abelian Berry curvatures. With these preparations, in Sec.~\ref{semiCcrnt-1-3}, we devise a theory for MSCT including non-adiabatic effects for evaluating the transport current within the phase-space framework.

\subsection{Semiclassical transport theory and the PSF}

\label{semiCcrnt-1-1}

\subsubsection{Fundamental elements of classical PSF}
\label{semiCcrnt-1-1-1}

In the classical analysis of the particle transport, we rely on the picture of a mass of fluid distributed in the phase space, so-called the PSF. The PSF is characterised by two ingredients. One ingredient is the mass distribution $f\left(\boldsymbol{x},\hbar\boldsymbol{k}\right)$ and the other is the fluid's velocity distribution $\boldsymbol{\mathcal{V}}\left(\boldsymbol{x},\hbar\boldsymbol{k}\right)$. The former specifies the amount of the PSF's mass occupying an elemental area of the phase-space centred around $\left(\boldsymbol{x},\hbar\boldsymbol{k}\right)$. The latter $\boldsymbol{\mathcal{V}}\left(\boldsymbol{x},\hbar\boldsymbol{k}\right)=\left(\dot{\boldsymbol{x}},\hbar\dot{\boldsymbol{k}}\right)$ has a position component $\dot{\boldsymbol{x}}$ and a momentum component $\hbar\dot{\boldsymbol{k}}$. The PSF's velocity are specified by the EOM for a single particle. The particle transport current density $\boldsymbol{J}\left(\boldsymbol{x}\right)$ at real-space
position $\boldsymbol{x}$ is given by
\begin{equation}
\boldsymbol{J}^{\text{cl.}}\left(\boldsymbol{x}\right)=\int \text{d}\boldsymbol{k}f\left(\boldsymbol{x},\hbar\boldsymbol{k}\right)\boldsymbol{v}\left(\boldsymbol{x},\hbar\boldsymbol{k}\right),\label{Bltzeq-current-def}
\end{equation} where $\boldsymbol{v}\left(\boldsymbol{x},\hbar\boldsymbol{k}\right)=\dot{\boldsymbol{x}}$ is built from the single-particle dynamics.

There is an important property of the classical transport theory that is to be inherited to a semiclassical construction which enables the description of the transport current for an electron gas. This property is that the fluid's mass distribution $f\left(\boldsymbol{x},\hbar\boldsymbol{k}\right)$ and velocity distribution $\boldsymbol{\mathcal{V}}\left(\boldsymbol{x},\hbar\boldsymbol{k}\right)$ are both completely determined by the phase-space coordinate $\left(\boldsymbol{x},\hbar\boldsymbol{k}\right)$. The former is an inherited part of the definition of a phase space. The latter is ensured by the nature of the classical Hamiltonian dynamics. This property is called the phase-space locality.

In summary, there are two fundamental elements in a classical transport theory. (i): The single-particle's EOM satisfies the phase-space locality. (ii): There exists a non-ambiguous distinction between the PSF's mass and velocity distributions. The velocity distribution is directly based on the scattering-free single-particle dynamics while the scattering-induced effects are included only through the mass distribution $f\left(\boldsymbol{x},\hbar\boldsymbol{k}\right)$.

\subsubsection{The correspondence between PSF and SSCT}
\label{semiCcrnt-1-1-2}

Here we discuss how the above fundamental elements of classical transport theory are preserved in the SSCT.

(i): The restriction of having only a single band in the active manifold lets the SC-EOM, Eq.~(\ref{phsvelo-sep-2-AB}), exhibit the phase-space locality. However, the phase-space locality is not guaranteed in Eq.~(\ref{phsvelo-sep-2}) (see discussions in Sec.~\ref{semiCcrnt-1-2}). Henceforth, a sufficiently small external field that enables one to focus on a single band at a time is vital to the phase-space locality.

(ii): Being able to focus at a time on a single-band manifold, indexed by its band $n$, also furnishes the specification of the PSF's mass for that band by associating it with an equilibrium Fermi-Dirac distribution $f^{0}_{n}$ and a non-equilibrium deviation $\delta f^{}_{n}$, namely,
\begin{subequations}
\label{non-eq-ad}
\begin{align}
f_{n}=f^{0}_{n}+\delta f^{}_{n},\label{non-eq-f}
\end{align}  in which
\begin{align}
f^{0}_{n}=\frac{1}{e^{\left(\varepsilon_{n}-\mu\right)/k_{B}T}+1},\label{eq-f0}
\end{align} with the chemical potential $\mu$, the temperature $T$ and Boltzmann constant $k_{B}$. Under the widely applied relaxation-time approximation~\cite{Chang951348,Chang967010,Deyo091917v1,Moore10026805,Sodemann15216806} the deviation from equilibrium reads
\begin{align}
\delta f^{}_{n}=-\tau\left[\frac{\partial f^{0}_{n}}{\partial\boldsymbol{x}}\cdot\dot{\boldsymbol{x}}_{n}
+\frac{\partial f^{0}_{n}}{\partial\boldsymbol{k}}\cdot\dot{\boldsymbol{k}}_{n}\right],\label{non-eq-1}
\end{align} where $\tau$ is the scattering time.
\end{subequations} Here $\left(\dot{\boldsymbol{x}}_{n},\hbar\dot{\boldsymbol{k}}_{n}\right)$ is given by Eq.~(\ref{phsvelo-sep-2-AB}) corresponding to the PSF's velocity which is decoupled from the scattering effects. The scattering effects are incorporated only in $\delta f^{}_{n}$. Therefore the SSCT expressed by Eqs.~(\ref{phsvelo-sep-2-AB}) and (\ref{non-eq-ad}) satisfies the property that the PSF's velocity and mass are well separated notions.

For the transport current, the classical formula Eq.~(\ref{Bltzeq-current-def}) is turned into a semiclassical one by replacing the term $f\boldsymbol{v}$ with the summation of contributions from all single-band manifolds, namely,
\begin{align}
\boldsymbol{J}^{\text{Ab}}\left(\boldsymbol{x}\right)=-e\sum_{n}\int\text{d}\boldsymbol{k}
f_{n}\dot{\boldsymbol{x}}_{n},\label{Bltzeq-2}
\end{align} which is straightforwardly analogue to Eq.~(\ref{Bltzeq-current-def}). The superscript Ab for the current reminds of appearance of Abelian Berry curvatures in $\dot{\boldsymbol{x}}_{n}$.

Since one focuses only on one band at a time, the band index is often omitted as a convention pointed out in the standard textbook.\cite{Ashcroft76book}
The results Eqs.~(\ref{non-eq-ad}) and (\ref{Bltzeq-2}) are well established in the literatures\cite{Ashcroft76book,Chang951348,Chang967010,Xiao101959} and we shall reproduce them from the more general developments in Sec.~\ref{semiCcrnt-1-3}.

\subsection{Characters of MSCT}

\label{semiCcrnt-1-2}

\subsubsection{The general absence of the phase-space locality}
\label{semiCcrnt-1-2-1}

For the case of multiple-band manifolds, regardless if these bands are degenerate or not, the single wavepacket SC-EOM for its centre-of-mass is given by Eq.~(\ref{phsvelo-sep-2}) which requires the explicit knowledge of $\bar{\boldsymbol{\eta}}_{a}$ governed by Eq.~(\ref{eta-d-lgg-1}). The formal solution to Eq.~(\ref{eta-d-lgg-1}) reads $\bar{\boldsymbol{\eta}}_{a}\left(t\right)=\hat{T}\exp\left\{ -i\int_{t_0}^{t}\text{d}\tau%
\bar{\mathcal{H}}_{a}\left(\boldsymbol{x}\left(\tau\right),\boldsymbol{k}\left(\tau\right)
\right) \right\} \bar{\boldsymbol{\eta}}_{a}\left(t_0\right)$. The time integral $\int_{t_0}^{t}\text{d}\tau$ says that $\bar{\boldsymbol{\eta}}_{a}\left(t\right)$ depends not only on $\left(\boldsymbol{x}\left(t\right),\boldsymbol{k}\left(t\right)\right)$ but also on the passed trajectories of $\left(\boldsymbol{x}\left(\tau\right),\boldsymbol{k}\left(\tau\right)\right)$ for $\tau$ in the time interval between some initial time $t_0$ and $t$. Therefore, the centre-of-mass velocity $\dot{\boldsymbol{x}}$ according to Eq.~(\ref{phsvelo-sep-2}), which evaluates on $\bar{\boldsymbol{\eta}}_{a}\left(t\right)$, in general also depends on these passed trajectories of the centre-of-mass and cannot be a pure function of its present coordinate in the phase space. In other words, the full coherent dynamics of band pseudospin by Eq.~(\ref{eta-d-lgg-1}) renders the loss of the phase-space locality for the centre-of-mass by Eq.~(\ref{phsvelo-sep-2}).

\subsubsection{The mass and velocity are not two distinct properties}
\label{semiCcrnt-1-2-2}
For an ensemble of wavepackets, several wavepackets can have the same centre-of-mass while occupying different band states. The most general way of expressing such occupation configuration with the centre-of-mass at $\left(\boldsymbol{x},\boldsymbol{k}\right)$ is via a density operator $\rho\left(\boldsymbol{x},\boldsymbol{k}\right)$.

We introduce the creation (annihilation) operator $a_{m}^{\dagger}$ ($a_{m}$) that
creates (annihilates) an electron on the bare band $m$ carrying the momentum $\hbar\boldsymbol{k}$ with the wavepacket centre $\boldsymbol{x}$.  With $\rho$ given, the total occupation number at that phase-space point is $\bar{N}=\sum_{m}\text{tr}\left(\rho a_{m}^{\dagger}a_{m}\right)$. The sum of all wavepackets' velocity $\left\langle\dot{\boldsymbol{x}}\right\rangle$ and the force $\hbar\left\langle\dot{\boldsymbol{k}}\right\rangle$ then read
\begin{subequations}
\label{phsvelo-M-def-0}
\begin{equation}
\left\langle\dot{\boldsymbol{x}}\right\rangle=\text{tr}\left(\rho\hat{\boldsymbol{V}}_{M}\right) ,\label{phsvelo-pos-M-def-0}
\end{equation}
 and
\begin{equation}
\hbar\left\langle\dot{\boldsymbol{k}}\right\rangle=(-e)\bar{N}\boldsymbol{E}+\boldsymbol{F}^{R},\label{phsvelo-mom-M-def-0}
\end{equation} with
\begin{equation}
\boldsymbol{F}^{R}=\text{tr}\left(\rho\hat{\boldsymbol{F}}_{M}\right),\label{phsvelo-force-M-def-0}
\end{equation}
 \end{subequations} in which,
\begin{equation}
\hat{\boldsymbol{V}}_{M}=\sum_{m,l}\left\langle \bar{u}_{m}\left\vert \frac{\partial\mathcal{H}^{c}}{\partial\hbar\boldsymbol{k}}\right\vert \bar{u}_{l}\right\rangle a_{m}^{\dagger}a_{l},\label{velo-opt-manypt-def-0}
\end{equation}
 and
\begin{equation}
\hat{\boldsymbol{F}}_{M}=-\sum_{m,l}\left\langle \bar{u}_{m}\left\vert \frac{\partial\mathcal{H}^{c}}{\partial\boldsymbol{x}}\right\vert \bar{u}_{l}\right\rangle a_{m}^{\dagger}a_{l},\label{force-opt-manypt-def-0}
\end{equation}
 are respectively the velocity and the force operators extended
to the second quantisation form. Here the trace operator, $\text{tr}\left(\cdot\right)$, is within the band space.

In the language of a PSF, the fluid then has a velocity which has position components given by Eq.~(\ref{phsvelo-pos-M-def-0}) and momentum components by Eq.~(\ref{phsvelo-mom-M-def-0}).
The carrier transport current density then reads
\begin{align}
\boldsymbol{J}^{}\left(\boldsymbol{x}\right)=-e\int\text{d}\boldsymbol{k}\left\langle\dot{\boldsymbol{x}}\right\rangle.\label{Bltzeq-3}
\end{align} In SSCT (Eqs.~(\ref{non-eq-ad}) and (\ref{Bltzeq-2})), the band occupation $f_{n}$ is analogue to the PSF's mass. However, for MSCT with Eq.~(\ref{Bltzeq-3}), the carrier distribution with respect to the bands (PSF's mass) has already been taken into account in the definition of $\left\langle\dot{\boldsymbol{x}}\right\rangle$ (PSF's velocity) given by Eq.~(\ref{phsvelo-M-def-0}) through $\rho$. The clear distinction between the two properties, mass and velocity, of a PSF, thus exits only for SSCT but not for MSCT.

%Nevertheless, since the description by Eqs.~(\ref{phsvelo-M-def-0}) through (\ref{Bltzeq-3}) is general, it should cover the SAB limit (see later discussions).

\subsubsection{Subtlety in extending SSCT to MSCT}
\label{semiCcrnt-1-2-3}

A multiple-band manifold with $N_{a}$ bands can be filled by $N$ wavepackets with $N\le N_{a}$. The $j$th wavepacket has the velocity $\dot{\boldsymbol{x}}_{j}$. The sum of $N$ wavepackets' velocity is $\left\langle\dot{\boldsymbol{x}}\right\rangle=\sum_{j=1}^{N}\dot{\boldsymbol{x}}_{j}$. For each wavepacket, one then needs to evolve Eq.~(\ref{phsvelo-sep-2}) to determine $\dot{\boldsymbol{x}}_{j}$ which requires the knowledge of the band pseudospin from Eq.~(\ref{eta-d-lgg-1}). This complication is not present for SSCT. Even if the velocity of each wavepacket is obtained, the calculation of the current by Eq.~(\ref{Bltzeq-3}) is still hindered by the loss of phase-space locality discussed in Sec.~\ref{semiCcrnt-1-2-1}.

The situation can be much simplified under the following restrictions. We first assume that all the $N_{a}$ active bands are degenerate over the whole Brillouin zone (BZ) and fully occupied $N=N_{a}$. Under these conditions, Eq.~(\ref{Bltzeq-3}) upon the substitution of $\left\langle\dot{\boldsymbol{x}}\right\rangle=\sum_{j=1}^{N_{a}}\dot{\boldsymbol{x}}_{j}$, where $\dot{\boldsymbol{x}}_{j}$ is found by Eq.~(\ref{phsvelo-sep-2}) for the $j$th wavepacket, then reads,~\cite{note-3}
\begin{align}
&J_{\alpha}^{\text{nAb}}\left(\boldsymbol{x}\right)=e\!\!\int\!\!\!\text{d}\boldsymbol{k}\sum_{j=1}^{N_{a}}\sum_{\beta}
\Bigg[\left\langle\mathcal{F}^{k_{\alpha}x_{\beta}}\right\rangle_{a}
\left\langle\frac{\partial{\mathcal{H}^{c}_{a}}}{\partial{k}_{\beta}}\right\rangle_{a}
\nonumber\\
&
-\left\langle\mathcal{F}^{k_{\alpha}k_{\beta}}\right\rangle_{a}
\left(\frac{e}{\hbar}E_{\beta}+
\left\langle\frac{\partial{\mathcal{H}^{c}_{a}}}{\partial{x}_{\beta}}\right\rangle_{a}
\right)
\Bigg]_{j}.
\label{nAB-anomcrnt-0}
\end{align} The superscript nAB has been added to the current for indicating the appearance of non-Abelian Berry curvatures. Eq.~(\ref{nAB-anomcrnt-0}) can be further simplified by removing the spatial textures, namely, $\mathcal{F}^{k_{\alpha}x_{\beta}}=0$ and $\left\langle\partial{\mathcal{H}^{c}_{a}}/\partial{x}_{\beta}\right\rangle_{a}=0$, leading to
\begin{equation}
J_{\alpha}^{\text{nAb}}=-\frac{e^2}{\hbar}\sum_{\beta}\int d\boldsymbol{k}\text{Tr}\left(\mathcal{F}^{ k_{\alpha}k_{\beta}}\right)E_{\beta},\label{anom-crnt-Ad-nAB-0}
\end{equation} which is a known expression for the Hall current in terms of the trace of the $N_{a}\times N_{a}$ non-Abelian Berry
curvature matrix.\cite{Shindou05399} Here $\boldsymbol{J}_{}^{\text{nAb}}$ does not depend on $\boldsymbol{x}$ since spatial variations have been removed. Eq.~(\ref{anom-crnt-Ad-nAB-0}) will be reproduced in Sec.~\ref{semiCcrnt-1-3}.

If a relevant multiple-band manifold is only partially filled (as exemplified in the third column of Fig.~\ref{schematics}), where non-adiabatic dynamics matters, then straightforward extension of SSCT to MSCT does not help to evaluate  Eq.~(\ref{Bltzeq-3}) in general. Nevertheless, the formulation of Eqs.~(\ref{phsvelo-M-def-0}) and (\ref{Bltzeq-3}) indicates that as long as one is able to obtain a proper density matrix $\rho$, the explicit evaluation of the semiclassical current expression using Eq.~(\ref{Bltzeq-3}) is still feasible. In fact, finding the density matrix suitable for an electron gas in transport scenarios is the mission targeted by the quantum kinetic theory.\cite{Vasko05book,Culcer17035106} Instead of being fully quantum with involved microscopic details, the present work aims to provide a phenomenological shortcut to a steady-state density matrix for semiclassically computing the transport current.

\subsection{A non-adiabatic semiclassical transport theory}
\label{semiCcrnt-1-3}

\subsubsection{the decoherence and the hybridised bands}
\label{decoh-pic}

The steady state described by Eq.~(\ref{non-eq-ad}) as a statistical mixture of populations on a number of  bands adiabatically-treated in the framework of SSCT has implied the absence of the inter-band coherence. This has important implication for the steady state in general. In a companion work,\cite{Tu19} we have shown that a finite electric field can coherently couple the two branches of a Dirac cone as a non-perturbation effect and the fully coherent band-pseudospin dynamics does not result in a steady-state current. Henceforth, for obtaining the steady-state current in the non-adiabatic regime, the decoherence should be manifested.

Here we briefly recall Ref.~[\onlinecite{Tu19}] for the results of such decoherence.
We start from the EOM for inter-band coherence, Eq.~(\ref{effSCE-1}), and add a noise term to $\bar{\mathcal{H}}$, the moving-frame effective Hamiltonian given by Eq.~(\ref{mvfrm-H-0}). By applying the standard Born-Markov decoherence theory for solving the steady-state density matrix for the band state, we find that the steady-state density matrix is diagonal in the eigenbasis of $\bar{\mathcal{H}}$, referred here as the hybridised bands since they are hybridisation of the original bands. The verification of using the hybridised bands as the basis for decoherence will be further discussed later. Here we first have to find these hybridised bands explicitly under the influence of spatial textures.

We denote the hybridised bands by $\left\vert\frak{u}_{i}\right\rangle$, namely,
\begin{align}
\bar{\mathcal{H}}\left\vert\frak{u}_{i}\right\rangle=\mathcal{E}_{i}\left\vert\frak{u}_{i}\right\rangle,\label{hybbs-2bds}
\end{align} where $i$ indexes a hybridised band with the corresponding energy $\mathcal{E}_{i}$. Eq.~(\ref{hybbs-2bds}) is for full bands and we will discuss later about focusing on one manifold.
With the spatial textures, the off-diagonal hybridisation coupling $\bar{\mathcal{H}}_{n,m}=V_{n,m}$ for $n\ne{m}$ given by Eq.~(\ref{hyb-coup}) involves the unknown band state because the determination of
the centre-of-mass $\left(\dot{\boldsymbol{x}}_{},\dot{\boldsymbol{k}}\right)$ in $V_{n,m}$ relies on it (see Eqs.~(\ref{velo-nq-2}),(\ref{force-nq-2}) and (\ref{force-nq-2g})). The eigenstates of $\bar{\mathcal{H}}$ thus depend on the band occupations determined by the decoherence, bringing in additional complications for self-consistently finding the wavefunctions of the hybridised bands. Indeed, the present description of the spatial textures relies on the local view which is only available when the spatial variation is very smooth. Therefore, we ignore the contributions of spatial variations to inter-band transitions and approximate $V_{n,m}$ by
\begin{align}
V_{n,m}\approx-
\left[\bar{\mathcal{R}}_{\boldsymbol{k}_{}}\right]_{n,m}\cdot\left(-e\boldsymbol{E}\right),
\label{Vnm-loc}
\end{align}
 concentrating on the non-adiabatic effect due to the electric field only. Since the local Hamiltonian, $\mathcal{H}^{c}$, depends on  $\boldsymbol{x}$, then the $\boldsymbol{x}$-dependence is retained in its eigenstate $\left\vert\bar{u}_{n}\right\rangle$ and
 $\left[\bar{\mathcal{R}}_{\boldsymbol{k}_{}}\right]_{n,m}=\left\langle\bar{u}_{n}\left\vert i\partial \bar{u}_{m}/\partial\boldsymbol{k}\right.\right\rangle$, the approximation Eq.~(\ref{Vnm-loc}) still keeps the dependence on $\boldsymbol{x}$ of $V_{n,m}$ .

\subsubsection{Formulating the current for MSCT}
\label{crnt-MSCT}

For SSCT, Eq.~(\ref{non-eq-ad}), obtained by the standard kinetic theory,~\cite{Ashcroft76book} is a statistical mixture among the original bands. Analogously in MSCT, with the electric field hybridising the original bands and the noise eliminating coherence between the hybridised bands, we have a statistical mixture among the hybridised bands, namely, \begin{align}
\text{tr}\left(\rho c^{\dagger}_{i}c_{j}\right)=\delta_{i,j}g^{}_{i},
\label{nad-dsmtx}
\end{align} where $c^{\dagger}_{i}$ creates an electron on the hybridised band $i$. The explicit form of occupation number $g_{i}$ is found through a route similar to the kinetic theory for SSCT  (see Appendix for details) and is given by
\begin{subequations}
\label{non-eq-nad}
\begin{align}
g_{i}=g^{0}_{i}+\delta g^{}_{i},\label{non-eq-g}
\end{align} with
\begin{align}
g^{0}_{i}=\frac{1}{e^{\left(\mathcal{E}_{i}-\mu\right)/k_{B}T}+1},
\label{g0eq}
\end{align} and
\begin{align}
\delta g^{}_{i}=-\tau\left[
\frac{\partial g^{0}_{i}}{\partial\boldsymbol{x}}\cdot\tilde{\boldsymbol{v}}_{i}
+\frac{\partial g^{0}_{i}}{\partial\hbar\boldsymbol{k}}\cdot\left(-e\boldsymbol{E}+\tilde{\boldsymbol{F}}_{i}\right)
\right],\label{non-eq-dg}
\end{align}
\end{subequations} in which
\begin{subequations}
\label{veloeq-0}
\begin{equation}
\tilde{\boldsymbol{v}}_{i}=\left\langle \frak{u}_{i}\left\vert \frac{\partial\mathcal{H}^{c}}{\partial\hbar\boldsymbol{k}}\right\vert \frak{u}_{i}\right\rangle
\label{eqvelo-x-i}
\end{equation}
 and
\begin{equation}
\tilde{\boldsymbol{F}}_{i}=-\left\langle \frak{u}_{i}\left\vert \frac{\partial\mathcal{H}^{c}}{\partial\boldsymbol{x}}\right\vert \frak{u}_{i}\right\rangle.
\label{eqvelo-k-i}
\end{equation}
\end{subequations}
Substituting Eq.~(\ref{nad-dsmtx}) into Eq.~(\ref{phsvelo-M-def-0}) with the aids of Eq.~(\ref{veloeq-0}), one obtains,
\begin{align}
\label{coll-velo-x-1}
\left\langle\dot{\boldsymbol{x}}\right\rangle=\sum_{i}g^{}_{i}\tilde{\boldsymbol{v}}_{i},
\end{align} which can be further substituted into Eq.~(\ref{Bltzeq-3}) for the final computation of the steady-state transport current as
\begin{align}
\boldsymbol{J}^{}\left(\boldsymbol{x}\right)=-e\sum_{i}\int\text{d}\boldsymbol{k}g^{}_{i}\tilde{\boldsymbol{v}}_{i}\label{Bltzeq-4}.
\end{align} Given a chemical potential $\mu$ and a temperature $T$, the semiclassical formulae Eqs.~(\ref{non-eq-nad}) to (\ref{Bltzeq-4}) enable the calculation of the transport currents for multiple bands without the ambiguity discussed in Sec.~\ref{semiCcrnt-1-2}.

Since Eq.~(\ref{hybbs-2bds}) behind Eqs.~(\ref{non-eq-nad}) to (\ref{Bltzeq-4}) is for the full-band description, we impose the separation of the full bands into different manifolds introduced in Sec.~\ref{wavpck-dys-2}. $\bar{\mathcal{H}}_{a}$, the projection of $\bar{\mathcal{H}}$ in Eq.~(\ref{mvfrm-H-0}) on the active manifold, has eigenstates $\left\vert\frak{u}_{i}^{0}\right\rangle$ and eigenenergies $\mathcal{E}_{i}^{0}$, namely,
\begin{align}
\bar{\mathcal{H}}_{a}\left\vert\frak{u}_{i}^{0}\right\rangle
=\mathcal{E}_{i}^{0}\left\vert\frak{u}_{i}^{0}\right\rangle.\label{hybbs-abds}
\end{align} Note that $\left\vert\frak{u}_{i}^{0}\right\rangle$ contains the electric field to all orders. Given the remoteness of the $r$-bands, the hybridised bands as the eigenstates in Eq.~(\ref{hybbs-2bds}) are then found by treating $a$-$r$ couplings, $V_{m,l}$ with $m\in{a}$ and $l\in{r}$ in $\bar{\mathcal{H}}$ of Eq.~(\ref{mvfrm-H-0}), as perturbations to the unperturbed eigenstates of $\left\vert\frak{u}_{i}^{0}\right\rangle$. That means, the state vector in Eq.~(\ref{veloeq-0}) is given by $\left\vert\frak{u}_{i}^{}\right\rangle=\left\vert\frak{u}_{i}^{0}\right\rangle+\left\vert\delta\frak{u}_{i}^{}\right\rangle$, where the perturbation correction reads $\left\vert\delta\frak{u}_{i}^{}\right\rangle=\sum_{l\in r}\sum_{m\in a}\frac{-V_{l,m}\bar{\eta}^{(i)}_{m} }{\mathcal{E}_{i}^{0}-\varepsilon_{l}}\left\vert\bar{u}_{l}\right\rangle$, where $\bar{\eta}^{(i)}_{m}=\left\langle\bar{u}_{m}\left\vert\frak{u}^{0}_{i}\right.\right\rangle$ is found from the diagonalisation of Eq.~(\ref{hybbs-abds}) with the approximation Eq.~(\ref{Vnm-loc}). Substituting this $\left\vert\frak{u}_{i}^{}\right\rangle$ into Eq.~(\ref{veloeq-0}), we obtain
\begin{subequations}
\label{veloeq-1}
\begin{align}
\tilde{{v}}_{i\alpha}=\tilde{{v}}_{i\alpha}^{0}
+\frac{e}{\hbar}\sum_{\beta}\left\langle\mathcal{F}^{k_{\alpha}k_{\beta}}\right\rangle_{a}^{(i)}
E_{\beta},
\label{veloeq-1-x}
\end{align} and
\begin{align}
\tilde{{F}}_{i\alpha}=\tilde{{F}}_{i\alpha}^{0}
-e\sum_{\beta}\left\langle\mathcal{F}^{x_{\alpha}k_{\beta}}\right\rangle_{a}^{(i)}
E_{\beta},
\label{veloeq-1-k}
\end{align} where
\begin{align}
\tilde{{v}}_{i\alpha}^{0}=\left\langle \frak{u}_{i}^{0}\left\vert \frac{\partial\mathcal{H}^{c}_{a}}{\partial\hbar{k}_{\alpha}}\right\vert\frak{u}_{i}^{0}\right\rangle,~
\tilde{{F}}_{i\alpha}^{0}=-\left\langle\frak{u}_{i}^{0}\left\vert \frac{\partial\mathcal{H}^{c}_{a}}{\partial{x}_{\alpha}}\right\vert\frak{u}_{i}^{0}\right\rangle
\end{align}
\end{subequations} for the $\alpha$th component of $\tilde{\boldsymbol{v}}_{i}$ and $\tilde{\boldsymbol{F}}_{i}$ respectively. Here,
\begin{align}
\left\langle\mathcal{F}^{\lambda_{\alpha}k_{\beta}}\right\rangle_{a}^{(i)}=\sum_{n,m\in{a}}\left(\bar{\eta}^{(i)}_{n}\right)^{*}
\mathcal{F}^{\lambda_{\alpha}k_{\beta}}_{n,m}\bar{\eta}^{(i)}_{m}
\end{align} with $\lambda_{\alpha}$ taken to be $x_{\alpha}$ or $k_{\alpha}$.

We now discuss the relation between the MSCT expressed by Eqs.~(\ref{non-eq-nad}) through (\ref{Bltzeq-4}) and the known result Eq.~(\ref{anom-crnt-Ad-nAB-0}) (obtained without spatial textures, see Sec.~\ref{semiCcrnt-1-2-3}). We leave the discussions for spatial textures and comparison with SSCT to Sec.~\ref{effrstext}. Within a multiple-band manifold, the first term of Eq.~(\ref{veloeq-1-x}) can be decomposed as $\tilde{\boldsymbol{v}}_{i}^{0}=\tilde{\boldsymbol{v}}^{0}_{b,i}+\tilde{\boldsymbol{v}}^{0}_{h,i}$, where $\tilde{\boldsymbol{v}}_{b,i}^{0}=\sum_{n\in{a}}\left\vert \bar{\eta}^{(i)}_{n}\right\vert ^{2}\partial\varepsilon_{n}/\partial(\hbar\boldsymbol{k})$ and
$\tilde{\boldsymbol{v}}^{0}_{h,i}=\sum_{n\in{a}}\!\!\sum_{m\ne{n}\in{a}}\left(\bar{\eta}^{(i)}_{n}\right)^{*}
\left\langle \bar{u}_{n}\left\vert\partial\mathcal{H}^{c}_{a}/\partial(\hbar\boldsymbol{k})\right\vert \bar{u}_{m}\right\rangle\bar{\eta}^{(i)}_{m}$. The result of Eq.~(\ref{anom-crnt-Ad-nAB-0}) in Sec.~\ref{semiCcrnt-1-2} is obtained under the assumptions that the active bands are degenerate over the entire BZ and fully occupied. Under this degenerate condition, $\tilde{\boldsymbol{v}}^{0}_{h,i}=0$ by Eq.~(\ref{deg-vh}) and $\tilde{\boldsymbol{v}}^{0}_{i}=\tilde{\boldsymbol{v}}^{0}_{b,i}$. The full occupancy condition, $g_{i}=1$ for all $i\in{a}$, then gives $\int\text{d}\boldsymbol{k}g_{i}\tilde{\boldsymbol{v}}_{i}^{0}=0$ leaving the second term of Eq.~(\ref{veloeq-1-x}) to be the only non-vanishing contribution to $\boldsymbol{J}$ and turns Eq.~(\ref{Bltzeq-4}) to be Eq.~(\ref{anom-crnt-Ad-nAB-0}).

Note that by Eq.~(\ref{veloeq-0}), the second term of Eq.~(\ref{veloeq-1-x}) comes from a perturbation correction, as $\left\langle \frak{u}_{i}^{0}\left\vert \partial\mathcal{H}^{c}_{a}/\partial(\hbar\boldsymbol{k})\right\vert\delta\frak{u}_{i}^{}\right\rangle+\text{c.c.}$ If other manifolds are very remote from the active one such that the correction is negligible, then this second term of Eq.~(\ref{veloeq-1-x}) can be just ignored. In this case, the energy splitting among the original bands (and henceforth a nonzero anomalous velocity $\tilde{\boldsymbol{v}}^{0}_{h,i}\ne0$) becomes crucial to transport current. This is important when the active bands are only degenerate at isolated points but not the entire BZ. In an associated study,~\cite{Tu19} we considered a model of two bands given by a gapped Dirac cone of small gap size. The coupling to bands beyond the Dirac cone has been completely ignored, keeping only the intra-manifold hybridisation. We found that the current arising from the anomalous velocity $\tilde{\boldsymbol{v}}^{0}_{h,i}$ manifests the underlying non-adiabatic dynamics and conveys a non-perturbative nonlinear valley Hall effect, in the absence of the spatial textures. The present establishment Eqs.~(\ref{non-eq-nad}) through (\ref{Bltzeq-4}) includes also the possibility to take into account the spatial textures.

%Note that, the effects of non-adiabatic dynamics from non-degenerate bands within the active manifold can now be seen by removing from above the restriction of degeneracy. This leads to a current $\boldsymbol{J}=\sum_{i}\boldsymbol{J}^{(i)}_{h}+\boldsymbol{J}^{\text{nAb}}$, where the extra contribution $\boldsymbol{J}_{dh}$ from the anomalous velocity due to inter-band transitions within the active manifold reads,

%\begin{align}
%\boldsymbol{J}^{(i)}_{h}=-ie\int\!\!\text{d}\boldsymbol{k}\!\!\sum_{n\in{d}}\!\!\sum_{m\ne{n}\in{d}}
%\left\langle \bar{u}_{n}\left\vert \frac{\partial\mathcal{H}^{c}_{d}}{\partial\hbar\boldsymbol{k}}\right\vert \bar{u}_{m}\right\rangle.
%\end{align}

\subsubsection{The spatial textures}
\label{effrstext}

Here we discuss subtle issues related to the presence of spatial textures. We compare the spatial texture effects obtained from Eqs.~(\ref{non-eq-nad}) through (\ref{Bltzeq-4}) with that obtained by Eqs.~(\ref{non-eq-ad}) and (\ref{Bltzeq-2}) summarised in Sec.\ref{semiCcrnt-1-1-2} for SSCT.

Note that SSCT is arrived when the electric field is infinitesimal.
The bands are only weakly hybridised and the off-diagonals of Eq.~(\ref{mvfrm-H-0}) are treated as perturbation in the diagonalisation of $\bar{\mathcal{H}}$. Therefore, the original band index $n$ remains a good quantum number for labeling a weakly hybridised band in Eqs.~(\ref{non-eq-nad}) through (\ref{Bltzeq-4}). To the first order of the electric field, the band energies are $\mathcal{E}_{n}=\varepsilon_{n}$, leading to ${g}^{0}_{n}={f}^{0}_{n}$ in Eq.~(\ref{g0eq}). We then write Eq.~(\ref{non-eq-g}) as $g_{n}=f^{0}_{n}+\delta{g}_{n}$ so to be compared directly with Eq.~(\ref{non-eq-f}), $f_{n}=f^{0}_{n}+\delta{f}_{n}$. Explicitly, these non-equilibrium deviations read
\begin{align}
\delta{g}^{}_{n}=-\tau\frac{\partial{f}^{0}_{n}}{\partial\varepsilon_{n}}
\left\{-\boldsymbol{F}_{b,n}\cdot\tilde{\boldsymbol{v}}_{n}
+\boldsymbol{v}_{b,n}\cdot
\left(-e\boldsymbol{E}+\tilde{\boldsymbol{F}}_{n}\right)
\right\},\label{non-eq-dg-SAB}
\end{align}
\begin{align}
\delta{f}^{}_{n}=-\tau\frac{\partial{f}^{0}_{n}}{\partial\varepsilon_{n}}
\left\{-\boldsymbol{F}_{b,n}\cdot\dot{\boldsymbol{x}}_{n}
+\boldsymbol{v}_{b,n}\cdot
\hbar\dot{\boldsymbol{k}}_{n}
\right\},\label{non-eq-df-SAB}
\end{align} where $\boldsymbol{v}_{b,n}=\partial\varepsilon_{n}/\partial\hbar\boldsymbol{k}$ is the normal velocity of the band, and
\begin{align}
\boldsymbol{F}_{b,n}=-\frac{\partial\varepsilon_{n}}{\partial\boldsymbol{x}},
\label{frce-nrml}
\end{align}
is the force directly associated with the spatial variation of the dispersion. Comparing Eq.~(\ref{non-eq-dg-SAB}) with Eq.~(\ref{non-eq-df-SAB}), we see that $\tilde{\boldsymbol{v}}_{n}$ and $-e\boldsymbol{E}+\tilde{\boldsymbol{F}}_{n}$ play the roles comparable to $\dot{\boldsymbol{x}}_{n}$ and $\hbar\dot{\boldsymbol{k}}_{n}$ respectively. We will see that at $\boldsymbol{E}=0$, $\delta{g}^{}_{n}=0$ while $\delta{f}^{}_{n}\ne0$ due to terms of second order derivatives in $\boldsymbol{x}$.

In what follows, $\overleftrightarrow{\Omega}_{n}^{KX}$ denotes a matrix whose elements are defined by
$\left[\overleftrightarrow{\Omega}_{n}^{KX}\right]_{\alpha,\beta}=\Omega_{n}^{k_{\alpha}x_{\beta}}$ (see Eq.~(\ref{BCV-1}) ) with $\alpha,\beta$ indexing the spatial directions so $\overleftrightarrow{\Omega}_{n}^{KX}$ acts on vectors in real space.  Similar definitions are applied to $\overleftrightarrow{\Omega}^{XX}$, $\overleftrightarrow{\Omega}^{KK}$ and $\overleftrightarrow{\Omega}^{XK}$. Reducing the number of bands in the active manifold for Eq.~(\ref{veloeq-1}) to one, we have
\begin{subequations}
\label{Fn-weakhyb}
\begin{equation}
\tilde{\boldsymbol{v}}_{n}=\boldsymbol{v}_{b,n}+\boldsymbol{v}_{h,n},
\end{equation} and
\begin{equation}
\tilde{\boldsymbol{F}}_{n}=\boldsymbol{F}_{b,n}+\boldsymbol{F}_{h,n},
\end{equation} where
\begin{align}
\boldsymbol{v}_{h,n}=-\frac{e}{\hbar}\overleftrightarrow{\Omega}_{n}^{KK}\boldsymbol{E},
\end{align} is
the familiar anomalous velocity due to textures of Bloch bands in momentum space, and
\begin{equation}
\boldsymbol{F}_{h,n}=-e\overleftrightarrow{\Omega}_{n}^{XK}\boldsymbol{E},
\end{equation}
\end{subequations}
is termed the "anomalous" force with the Berry curvature $\overleftrightarrow{\Omega}_{n}^{XK}$ involved, discriminating itself from the "normal" force $\boldsymbol{F}_{b,n}$, Eq.~(\ref{frce-nrml}). Note that when $\boldsymbol{E}=0$, Eq.~(\ref{Fn-weakhyb}) reduces to $\tilde{\boldsymbol{v}}_{n}=\boldsymbol{v}_{b,n}$ and $\tilde{\boldsymbol{F}}_{n}=\boldsymbol{F}_{b,n}$. Consequently, $\delta{g}_{n}=0$ in Eq.~(\ref{non-eq-dg-SAB}) and with $\int\text{d}\boldsymbol{k}f^{0}_{n}\boldsymbol{v}_{b,n}=0$, Eq.~(\ref{Bltzeq-4}) gives $\boldsymbol{J}=0$ at $\boldsymbol{E}=0$. Eqs.~(\ref{non-eq-nad}) through (\ref{Bltzeq-4}) thus ensure zero current when no external field is applied.

% $\left\vert\tilde{u}_{i}\right\rangle\rightarrow\left\vert\bar{u}_{n}\right\rangle$.

Next we turn to Eq.~(\ref{non-eq-df-SAB}), which is extracted from Eq.~(\ref{non-eq-ad}), and consider $\dot{\boldsymbol{x}}_{n}$ and $\hbar\dot{\boldsymbol{k}}_{n}$ up to the same order of the small parameter $\epsilon$ (c.f. Eq.~(\ref{epsilon})). The solution to Eq.~(\ref{phsvelo-sep-2-AB}) for $\left(\dot{\boldsymbol{x}}_{n},\hbar\dot{\boldsymbol{k}}_{n}\right)$ reads
\begin{subequations}
\label{ad-xksol}
\begin{align}
\dot{\boldsymbol{x}}_{n}=\tilde{\boldsymbol{v}}_{n}+\delta\tilde{\boldsymbol{v}}_{n}
\label{ad-xksol-x}
\end{align} and
\begin{align}
\hbar\dot{\boldsymbol{k}}_{n}=-e\boldsymbol{E}+\tilde{\boldsymbol{F}}_{n}+\delta\tilde{\boldsymbol{F}}_{n}
\label{ad-xksol-k}
\end{align} where
\begin{align}
\delta\tilde{\boldsymbol{v}}_{n}=-\overleftrightarrow{\Omega}_{n}^{KX}\boldsymbol{v}_{b,n}
-\overleftrightarrow{\Omega}_{n}^{KK}\boldsymbol{F}_{b,n}/\hbar,
\label{ad-xksol-dx}
\end{align} and
\begin{align}
\delta\tilde{\boldsymbol{F}}_{n}=\hbar\overleftrightarrow{\Omega}_{n}^{XX}\boldsymbol{v}_{b,n}
+\overleftrightarrow{\Omega}_{n}^{XK}\boldsymbol{F}_{b,n}.
\label{ad-xksol-dk}
\end{align}
\end{subequations} Substituting Eq.~(\ref{ad-xksol}) into Eq.~(\ref{non-eq-df-SAB}), one obtains
\begin{subequations}
\label{ad-dev-1}
\begin{align}
\delta{f}_{n}=\delta{g}_{n}+\delta{f}^{\prime}_{n},
\end{align} where
\begin{align}
\delta{f}^{\prime}_{n}=-\tau\frac{\partial{f}^{0}_{n}}{\partial\varepsilon_{n}}\left(
-\boldsymbol{F}_{b,n}\cdot\delta\tilde{\boldsymbol{v}}_{n}+
\boldsymbol{v}_{b,n}\cdot\delta\tilde{\boldsymbol{F}}_{n}
\right).
\end{align}
\end{subequations} Without spatial textures, namely, $\overleftrightarrow{\Omega}_{n}^{KX}=\overleftrightarrow{\Omega}_{n}^{XX}=0$ and $\boldsymbol{F}_{b,n}=0$, then Eqs.~(\ref{ad-xksol}) and (\ref{ad-dev-1}) give $\delta\tilde{\boldsymbol{v}}_{n}=0$ and $\delta\tilde{\boldsymbol{F}}_{n}=0$ with $\dot{\boldsymbol{x}}_{n}=\tilde{\boldsymbol{v}}_{n}$, $\hbar\dot{\boldsymbol{k}}_{n}=-e\boldsymbol{E}+\tilde{\boldsymbol{F}}_{n}$ so $\delta{f}^{\prime}_{n}=0$ leading to $\delta{f}_{n}=\delta{g}_{n}$. Consequently, Eqs.~(\ref{non-eq-nad}) and (\ref{Bltzeq-4}) reduce to Eqs.~(\ref{non-eq-ad}) and (\ref{Bltzeq-2}).

%$\dot{\boldsymbol{x}}_{n}=\tilde{\boldsymbol{v}}_{n}$, $\hbar\dot{\boldsymbol{k}}_{n}=-e\boldsymbol{E}+\tilde{\boldsymbol{F}}_{n}$ and $\delta{f}_{n}=\delta{g}_{n}$.

With spatial textures under the smooth spatial variation approximation Eq.~(\ref{Vnm-loc}), $\tilde{\boldsymbol{v}}_{i}$ is zeroth order and $\tilde{\boldsymbol{F}}_{i}$ is first order in the derivative of $\boldsymbol{x}$ respectively by Eq.~(\ref{veloeq-0}). Therefore, continuing the discussion for single-band manifolds, $\delta{g}_{n}$ in Eq.~(\ref{non-eq-dg-SAB}) is first order in $\partial/\partial\boldsymbol{x}$.
Comparing Eq.~(\ref{ad-dev-1}) with Eq.~(\ref{non-eq-dg-SAB}), we find $\delta{f}_{n}=\delta{g}_{n}$ by discarding $\delta{f}^{\prime}_{n}$ which is second order in the derivative of $\boldsymbol{x}$. This corresponds to neglecting $\delta\tilde{\boldsymbol{v}}_{n}$ and $\delta\tilde{\boldsymbol{F}}_{n}$ in Eqs.~(\ref{ad-xksol-x}) and (\ref{ad-xksol-k}), recovering $\dot{\boldsymbol{x}}_{n}=\tilde{\boldsymbol{v}}_{n}$ and $\hbar\dot{\boldsymbol{k}}_{n}=-e\boldsymbol{E}+\tilde{\boldsymbol{F}}_{n}$. We thus conclude that at $\boldsymbol{E}=0$, Eq.~(\ref{non-eq-dg-SAB}) gives $\delta{g}_{n}=0$ and therefore also zero current (see the discussions below Eq.~(\ref{Fn-weakhyb})). Meanwhile, with $\boldsymbol{E}=0$, the conventional SSCT described by Eq.~(\ref{non-eq-df-SAB}) (rewritten as Eq.~(\ref{ad-dev-1})) gives $\delta{f}_{n}=\delta{f}^{\prime}_{n}\ne0$ anticipating a nonzero current due to a non-vanishing deviation from equilibrium $\delta{f}^{\prime}_{n}$ caused by spatial inhomogeneity, second order in the derivative of $\boldsymbol{x}$.

When $\boldsymbol{E}\ne0$,  the effect of spatial textures will be manifested. Plugging Eq.~(\ref{Fn-weakhyb}) in Eq.~(\ref{non-eq-dg-SAB}) gives,
\begin{subequations}
\label{ad-dev}
\begin{align}
\delta{g}_{n}=\delta{g}^{b}_{n}+\delta{g}^{h}_{n},
\end{align} where
\begin{align}
\delta{g}^{b}_{n}=-\tau\frac{\partial{f}^{0}_{n}}{\partial\varepsilon_{n}}
\boldsymbol{v}_{b,n}\cdot\left(-e\boldsymbol{E}\right)
\end{align}
comes from the "normal" contribution from the normal velocity $\boldsymbol{v}_{b,n}$ and the electric force $\left(-e\boldsymbol{E}\right)$
and
\begin{align}
\delta{g}^{h}_{n}=-\tau\frac{\partial{f}^{0}_{n}}{\partial\varepsilon_{n}}
\left\{
\boldsymbol{v}_{b,n}\cdot\boldsymbol{F}_{h,n}
+\boldsymbol{F}_{b,n}\cdot\boldsymbol{v}_{h,n}
\right\},
\end{align}
\end{subequations}
is the "anomalous" contribution involving the "anomalous" spatial force $\boldsymbol{F}_{h,n}$ and the familiar anomalous velocity $\boldsymbol{v}_{h,n}$. Notably, this anomalous contribution $\delta{g}^{h}_{n}$ is exclusively due to the spatial textures, i.e. requiring $\boldsymbol{F}_{h,n}\ne0$, $\boldsymbol{F}_{b,n}\ne0$.

Note that the approximation Eq.~(\ref{Vnm-loc}) for smooth spatial variations places no limitation on the magnitude of $\boldsymbol{E}$. Finite electric field is responsible for inducing non-adiabatic dynamics for partially filled multiple-band manifold. As discussed for Eq.~(\ref{hybbs-abds}), $\left\vert\frak{u}_{i}^{0}\right\rangle$ contains all orders of $\boldsymbol{E}$. This leads  $\tilde{\boldsymbol{v}}_{i}$ and $\tilde{\boldsymbol{F}}_{i}$ as well as $\delta{g}_{i}$ to be also all orders in $\boldsymbol{E}$, reflecting the non-adiabatic dynamics among active bands.

%The spatial variation of the current implies that in different locations with different band structures, the rate of change of the local charge density can be different and depend on the spatial variation of the local wavefunction, namely,  $\partial\left\vert\bar{u}_{n}\left(\boldsymbol{x},\boldsymbol{k}\right)\right\rangle/\partial\boldsymbol{x}$ through the term $\boldsymbol{\nabla}\cdot\boldsymbol{J}$.

\section{Summary}

In the first part, concerning the SC-EOM of a single wavepacket, we have shown that the
same set of EOM, Eqs.~(\ref{velo-nq-1}) and (\ref{force-nq-1}), obtained by the Lagrangian variational approach applied to a wavepacket ansatz, can be equally arrived by quantum evolution of the carrier's wavefunction by the local Hamiltonian from which the wavepacekt's centre-of-mass are obtained by quantum expectation values in form of Eqs.~(\ref{velo-nq-2}) and (\ref{force-nq-2}). This SC-EOM for a single wavepacket paves the way for discussing non-adiabatic transport for a gas of electrons as an ensemble of wavepacket.

In the second part, we expounded the difficulty of straightforward extending SSCT to MSCT. When there is a multiple-band manifold, the coherent superpposition among the several active bands makes the determination of the band state and therefore the velocity expectation value complicated. Due to the loss of phase-space locality of the wavepacket's velocity, characteristic to multiple-band manifolds, this also results in the incapacity of semiclassically evaluating the current. We circumvent this obstacle by explicating the loss of inter-band coherence within the kinetic theory. This leads to the establishment of Eqs.~(\ref{non-eq-nad}) to (\ref{Bltzeq-4}). By inspecting the spatial texture effects, we find that the present MSCT is suitable for smooth spatial variations and ensures that no current can appear unless an external electric field is applied. On the contrary, the known SSCT, Eqs.~(\ref{non-eq-ad}) and (\ref{Bltzeq-2}) has implied a finite current even in the absence of electric field, that is second order in the spatial derivative.

We now comment on the range of applicability of the MSCT described by Eqs.~(\ref{non-eq-nad}) to (\ref{Bltzeq-4}). First, although we are intended to the regime where the electric field is finite other than infinitesimal, the magnitude of the electric field should be limited to not push away the electron gas too far from an equilibrium. Otherwise, the kinetic construction with relaxation time approximation simply fails. Second, we have assumed that the scattering rate is independent of the band index. How to go beyond the relaxation time approximation and construct a kinetic theory for the single carrier's dynamics governed by Eqs.~(\ref{velo-nq-2}) and (\ref{force-nq-2}) is out of the scope of the present attempt. Albeit these limitations, given the wide uses of the semiclassical transport theory based on Eqs.~(\ref{non-eq-ad}) and (\ref{Bltzeq-2}), the above extension to Eqs.~(\ref{non-eq-nad}) to (\ref{Bltzeq-4}) with finite electric field should also find its place of applications in materials with small gaps as discussed in the introduction.
\label{Sec-con}

\begin{acknowledgements}
The work is supported by the Research Grants Council of Hong Kong (Grants No. HKU17306819 and No. C7036-17W), and the University of Hong Kong (Seed Funding for Strategic Interdisciplinary Research). We thank Hui-Yuan Zheng and Hongyi Yu for useful discussions.
\end{acknowledgements}

\appendix

\section{A kinetic theory for MSCT}

Here we closely follow the kinetic theory for SSCT~\cite{Ashcroft76book} and make proper modifications for MSCT to obtain Eq.~(\ref{non-eq-nad}).
For SSCT, the evolution of a wavepacket according to Eq.~(\ref{phsvelo-sep-2-AB}) does not lead to inter-band coherence. The decoherence issue does not appear in the kinetic theory leading to the known SSCT. However, for multiple-band manifolds, a wavepacket can evolve to coherent superpositions among bands. Therefore, to construct a proper kinetic theory in this case, one needs to take into account the decoherence explicitly. For the ease of reference, the kinetic theory for SSCT is reviewed in Sec.~\ref{kic-ad}. Extension of the kinetic theory from SSCT to MSCT is discussed in Sec.~\ref{kic-nad}.

\subsection{The building blocks of the kinetic theory}
\label{kic-ad}

The kinetic theory with the relaxation-time approximation giving rise to Eq.~(\ref{non-eq-ad}) is based on the following assumptions.
(i): The scattering within the electron gas maintains it at equilibrium. Given the set of bands indexed by $n$, a natural choice of the equilibrium distribution function is given by Eq.~(\ref{eq-f0}).
(ii): The scattering rate does not depend on the form of the distribution function.
(iii): The EOM governing the dynamics of a single carrier free from scattering is given by Eq.~(\ref{phsvelo-sep-2-AB}) so that an electron starting from one band cannot cause nonzero occupations on other bands.

We denote by $f_{n}(\boldsymbol{\lambda},t)$ the occupation on a band $n$ at at a phase-space point $\boldsymbol{\lambda}=\left(\boldsymbol{x},\hbar\boldsymbol{k}\right)$ at time $t$. The assumption (i) corresponds to decompose $f_{n}$ as Eq.~(\ref{non-eq-f}) in which $\delta{f}_{n}$ is to be derived from assumptions (ii) and (iii).  The differential, $\text{d}f_{n}(\boldsymbol{\lambda},t)$, is understood to be the number of electrons gained into the state, $\left(n,\boldsymbol{\lambda}\right)$. Based on these assumptions, the derivation of Eq.~(\ref{non-eq-1}) is divided into the following basic blocks (a)-(d).

(a): Combining the assumptions (i) and (ii), one deduces that the number of electrons loss from $\left(n,\boldsymbol{\lambda}\right)$ from an equilibrium distribution in an infinitesimal time interval $\text{d}t$ due to scattering should be compensated by the number of electrons gained into the same state, namely,
\begin{align}
\label{eq-scatt-fm}
\text{d}f_{n}\left(\boldsymbol{\lambda},t\right)=\frac{\text{d}t}{\tau_{n}\left(\boldsymbol{\lambda}\right)}f_{n}^{0}\left(\boldsymbol{\lambda}\right),
\end{align} in which $\tau_{n}\left(\boldsymbol{\lambda}\right)$ is the scattering time whose inverse gives the scattering rate.

(b): The deterministic evolution of Eq.~(\ref{phsvelo-sep-2-AB}) for a time interval from time $t^{\prime}$ to $t$ would bring the state $\left(n,\boldsymbol{\lambda}^{\prime}=\boldsymbol{\lambda}(t^{\prime})\right)$ to the state $\left(n,\boldsymbol{\lambda}=\boldsymbol{\lambda}(t)\right)$.
The electrons occupying the state $\left(n,\boldsymbol{\lambda}\right)$ at time $t$ are therefore contributed by the no-scattering fraction of those that had occupied $\left(n,\boldsymbol{\lambda}^{\prime}\right)$ at time $t^{\prime}<t$. Denoting by $P_{\left(n,\boldsymbol{\lambda}^{\prime}\right)}\left(t,t^{\prime}\right)$ as the probability that such evolution can actually be reached without scattering, the occupations at time $t$ are related to all prior times by
\begin{align}
\label{noscatt-stream}
f_{n}\left(\boldsymbol{\lambda},t\right)=
\int_{t^{\prime}=-\infty}^{t^{\prime}=t}\text{d}f_{n}\left(\boldsymbol{\lambda}^{\prime},t^{\prime}\right)
P_{\left(n,\boldsymbol{\lambda}^{\prime}\right)}\left(t,t^{\prime}\right)
\end{align}

(c): The general property of $P_{\left(n,\boldsymbol{\lambda}^{\prime}\right)}\left(t,t^{\prime}\right)$ is that the no-scattering probability for a time interval from $t^{\prime}$ to $t$ is smaller than that for a time interval from $t^{\prime}+\text{d}t^{\prime}$ to $t$ by a factor $\left(1-\text{d}t^{\prime}/\tau\left(\boldsymbol{\lambda}^{\prime}\right)\right)$. More explicitly, this reads
$P_{\left(n,\boldsymbol{\lambda}^{\prime}\right)}\left(t,t^{\prime}\right)
=P_{\left(n,\boldsymbol{\lambda}^{\prime}\right)}\left(t,t^{\prime}+\text{d}t^{\prime}\right)
\left(1-\text{d}t^{\prime}/\tau_{n}\left(\boldsymbol{\lambda}^{\prime}\right)\right)$ and consequently
\begin{align}
\label{noscatt-prob-ad-dv}
\frac{\partial}{\partial t^{\prime}}P_{\left(n,\boldsymbol{\lambda}^{\prime}\right)}\left(t,t^{\prime}\right)
=\frac{P_{\left(n,\boldsymbol{\lambda}^{\prime}\right)}\left(t,t^{\prime}\right)}
{\tau_{n}\left(\boldsymbol{\lambda}^{\prime}\right)}.
\end{align} Furthermore, this no-scattering probability also satisfies
\begin{subequations}
\label{noscatt-prob-ad-bd}
\begin{align}
P_{\left(n,\boldsymbol{\lambda}^{\prime}\right)}\left(t,t\right)=1,
\end{align} simply for that when $t=t^{\prime}$ there is no time for scattering to occur between $t^{\prime}$ and $t$ and
\begin{align}
P_{\left(n,\boldsymbol{\lambda}^{\prime}\right)}\left(t,-\infty\right)=0,
\end{align} for scattering definitely occurs in a long enough time interval.
\end{subequations}

(d): Combining Eqs.~(\ref{eq-scatt-fm}) and (\ref{noscatt-prob-ad-dv}) into Eq.~(\ref{noscatt-stream}) and performing the resulted integral over $\text{d}t^{\prime}$ via the integration by part with the boundary condition Eq.~(\ref{noscatt-prob-ad-bd}), we obtain an intermediate expression
\begin{align}
\label{interm-fn}
f_{n}\left(\boldsymbol{\lambda},t\right)=f_{n}^{0}\left(\boldsymbol{\lambda}\right)-
\int_{t^{\prime}=-\infty}^{t^{\prime}=t}\!\!\!\!\!\!\!\!\!\!\!\!\!\!\!
\text{d}t^{\prime}e^{-\left(t-t^{\prime}\right)/\tau_{n}\left(\boldsymbol{\lambda}^{\prime}\right)}
\left[\frac{\text{d}}{\text{d}t^{\prime}}f_{n}^{0}\left(\boldsymbol{\lambda}\left(t^{\prime}\right)\right)\right].
\end{align} Assuming further that $\tau_{n}$ is very short so the factor $e^{-\left(t-t^{\prime}\right)/\tau_{n}\left(\boldsymbol{\lambda}^{\prime}\right)}$ only contributes significantly when $t^{\prime}$ is near $t$, the integrand in Eq.~(\ref{interm-fn}) is then approximated by  $\tau_{n}\left(\boldsymbol{\lambda}(t^{\prime})\right)\approx\tau_{n}\left(\boldsymbol{\lambda}\right)$ and $\left[\frac{\text{d}}{\text{d}t^{\prime}}f_{n}^{0}\left(\boldsymbol{\lambda}\left(t^{\prime}\right)\right)\right]\approx
\left[\frac{\text{d}}{\text{d}t^{}}f_{n}^{0}\left(\boldsymbol{\lambda}\left(t^{}\right)\right)\right]$, and Eq.~(\ref{interm-fn}) reduces to $f_{n}\left(\boldsymbol{\lambda}\right)=f_{n}^{0}\left(\boldsymbol{\lambda}\right)-\tau
\left[\frac{\text{d}}{\text{d}t^{}}f_{n}^{0}\left(\boldsymbol{\lambda}\left(t^{}\right)\right)\right]$, writing simply $\tau=\tau_{n}\left(\boldsymbol{\lambda}\right)$, a precursor of Eq.~(\ref{non-eq-ad}). By further expressing $\left[\frac{\text{d}}{\text{d}t^{}}f_{n}^{0}\left(\boldsymbol{\lambda}\left(t^{}\right)\right)\right]=\partial{f_{n}^{0}}/\partial\boldsymbol{\lambda}
\cdot\dot{\boldsymbol{\lambda}}_{n}$, Eq.~(\ref{non-eq-ad}) is recovered. Applying $\partial/\partial{t}$ to both sides of Eq.~(\ref{interm-fn}) with the short scattering-time approximation, one consistently obtains the stationarity of $f_{n}\left(\boldsymbol{\lambda},t\right)$, namely, $\partial{f}_{n}\left(\boldsymbol{\lambda},t\right)/\partial{t}=0$ so $f_{n}\left(\boldsymbol{\lambda},t\right)=f_{n}(\boldsymbol{\lambda})$. We are thus led to the SSCT summarised in Sec.~\ref{semiCcrnt-1-1-2}.

\subsection{Constructing the kinetic theory in the non-adiabatic regime}
\label{kic-nad}

For multiple-band manifolds, given a density matrix $\rho(t)$ at time $t$, the probability of having a wavepacket with band state $\left\vert\chi\right\rangle$ and the centre-of-mass, $\boldsymbol{\lambda}=\left(\boldsymbol{x},\hbar\boldsymbol{k}\right)$, is given by $g_{\chi}\left(\boldsymbol{\lambda},t\right)=\text{tr}\left(\rho(t)
a^{\dagger}_{\chi}(\boldsymbol{\lambda})a_{\chi}(\boldsymbol{\lambda})\right)$, where $a^{\dagger}_{\chi}(\boldsymbol{\lambda})$ creates an electron wavepacket of such a state.

Similar to the construction for SSCT in Sec.~\ref{kic-ad}, here we build a kinetic theory in the non-adiabatic regime from the same assumptions (i) and (ii) as before, with the original bands replaced by the hybridized bands (eigenstates described in Eq.~(\ref{hybbs-2bds})). The assumption (i) amounts to have Eq.~(\ref{nad-dsmtx}) with $g_{i}$ decomposed as Eq.~(\ref{non-eq-g}), but leaving the deviation $\delta{g}_{i}$ unspecified, whose explicit form will be derived here. More crucially, the underlying scattering-free single carrier's SC-EOM are now given by Eqs.~(\ref{velo-nq-2}) and (\ref{force-nq-2}), instead of Eq.~(\ref{phsvelo-sep-2-AB}). This allows non-perturbative inter-band transitions that create inter-band coherence.

%With regard to this, scattering-free evolution guided by Eqs.~(\ref{velo-nq-2}) and (\ref{force-nq-2}) generally evolves an electron from a state specified by $\left(\chi^{\prime}=\chi(t^{\prime}),\boldsymbol{\lambda}^{\prime}=\boldsymbol{\lambda}(t^{\prime})\right)$ to a state $\left(\chi^{}=\chi(t^{}),\boldsymbol{\lambda}^{}=\boldsymbol{\lambda}(t^{})\right)$.

The loss of coherence among the hybridised bands is accounted as the followings. Using $\left\vert\frak{u}_{i}\right\rangle$'s as a complete basis set, we have $a^{\dagger}_{\chi}=\sum_{i}\left\langle\left.\frak{u}_{i}\right\vert \chi\right\rangle c^{\dagger}_{i}$ and henceforth, $g_{\chi}=\text{tr}\left(\rho
a^{\dagger}_{\chi}a_{\chi}\right)=\sum_{i,j}\left\langle\left.\frak{u}_{i}\right\vert \chi\right\rangle\text{tr}\left(\rho
c^{\dagger}_{i}c_{j}\right)\left\langle\left.\chi\right\vert\frak{u}_{j}\right\rangle$. Combining this with Eq.~(\ref{nad-dsmtx}), which expresses no coherence among hybridised bands, we are led to
\begin{align}
\label{no-coh-hyb}
g_{\chi}\left(\boldsymbol{\lambda},t\right)=\sum_{i}\left\vert \left\langle\left.\frak{u}_{i}\right\vert \chi\right\rangle \right\vert ^{2}g_{i}\left(\boldsymbol{\lambda},t\right).
\end{align}  We will see that the kinetic theory for MSCT introduced here is in one-to-one correspondence to that of SSCT. The building blocks (a)-(d) in Sec.~\ref{kic-ad} become ($\text{a}^{\prime}$)-($\text{d}^{\prime}$) below, with the additional use of the decoherence consequence, Eq.~(\ref{no-coh-hyb}).

($\text{a}^{\prime}$): The equilibrium being unaltered by scattering can still be formulated similar to Eq.~(\ref{eq-scatt-fm}) by
\begin{align}
\label{eq-scatt-gm}
\text{d}g_{i}\left(\boldsymbol{\lambda},t\right)=\frac{\text{d}t}{\tau_{}\left(\boldsymbol{\lambda}\right)}g_{i}^{0}\left(\boldsymbol{\lambda}\right),
\end{align} where for simplicity we ignored the band dependence of the scattering rate.

($\text{b}^{\prime}$): The scattering-free evolution guided by Eqs.~(\ref{velo-nq-2}) and (\ref{force-nq-2}) generally evolves an electron from a state specified by $\left(\chi^{\prime}=\chi(t^{\prime}),\boldsymbol{\lambda}^{\prime}=\boldsymbol{\lambda}(t^{\prime})\right)$ to a state $\left(\chi^{}=\chi(t^{}),\boldsymbol{\lambda}^{}=\boldsymbol{\lambda}(t^{})\right)$.
The occupation of a state $\left(\chi^{},\boldsymbol{\lambda}^{}\right)$ at time $t$ is related to all prior times by
\begin{align}
\label{noscatt-stream-gm}
g_{\chi}\left(\boldsymbol{\lambda},t\right)=
\int_{t^{\prime}=-\infty}^{t^{\prime}=t}\text{d}g_{\chi^{\prime}}\left(\boldsymbol{\lambda}^{\prime},t^{\prime}\right)
P_{\left(\chi^{\prime},\boldsymbol{\lambda}^{\prime}\right)}\left(t,t^{\prime}\right),
\end{align} where $P_{\left(\chi^{\prime},\boldsymbol{\lambda}^{\prime}\right)}\left(t,t^{\prime}\right)$ stands for the probability that a carrier occupying the state $\left(\chi^{\prime},\boldsymbol{\lambda}^{\prime}\right)$ at time $t^{\prime}$ would not be scattered in the interval between $t^{\prime}$ and $t$.

($\text{c}^{\prime}$): The property that the scattering occurs with a probability proportional to the length of time interval considered is very general. Therefore, by the similarity to Eq.~(\ref{noscatt-prob-ad-dv}) we have
\begin{align}
\label{noscatt-prob-ad-dv-gm}
\frac{\partial}{\partial t^{\prime}}P_{\left(\chi^{\prime},\boldsymbol{\lambda}^{\prime}\right)}\left(t,t^{\prime}\right)
=\frac{P_{\left(\chi^{\prime},\boldsymbol{\lambda}^{\prime}\right)}\left(t,t^{\prime}\right)}
{\tau_{}\left(\boldsymbol{\lambda}^{\prime}\right)},
\end{align} and surely also
\begin{subequations}
\label{noscatt-prob-ad-bd-gm}
\begin{align}
P_{\left(\chi^{\prime},\boldsymbol{\lambda}^{\prime}\right)}\left(t,t\right)=1,
\end{align} and
\begin{align}
P_{\left(\chi^{\prime},\boldsymbol{\lambda}^{\prime}\right)}\left(t,-\infty\right)=0.
\end{align}
\end{subequations}

($\text{d}^{\prime}$): Similar to the steps prescribed in (d), substituting Eqs.~(\ref{eq-scatt-gm}) and (\ref{noscatt-prob-ad-dv-gm}) with the boundary conditions Eq.~(\ref{noscatt-prob-ad-bd-gm}) into Eq.~(\ref{noscatt-stream-gm}), the integration by part performed to the integral over $\text{d}t^{\prime}$ leads to an integral equation similar to Eq.~(\ref{interm-fn}). Now on the left-hand side of Eq.~(\ref{noscatt-stream-gm}), we let $\left\vert\chi\right\rangle=\left\vert\tilde{u}_{i}\right\rangle$ so ${g}_{\chi}=g_{i}$. On the right-hand side of Eq.~(\ref{noscatt-stream-gm}), we replace $g_{\chi^{\prime}}$ in the integrand by Eq.~(\ref{no-coh-hyb}). These lead to
\begin{align}
\label{interm-gm}
g_{i}\left(\boldsymbol{\lambda},t\right)&=g_{i}^{0}\left(\boldsymbol{\lambda}\right)
\nonumber\\&-\sum_{i^{\prime}}
\int_{t^{\prime}=-\infty}^{t^{\prime}=t}\!\!\!\!\!\!\!\!\!\!\!\!\!\!\!
\text{d}t^{\prime}e^{-\left(t-t^{\prime}\right)/\tau_{}\left(\boldsymbol{\lambda}^{\prime}\right)}
\left\vert \left\langle \left.\tilde{u}_{i^{\prime}}\right\vert \chi^{\prime}\right\rangle \right\vert ^{2}
\left[\frac{\text{d}}{\text{d}t^{\prime}}g_{i^{\prime}}^{0}\left(\boldsymbol{\lambda}\left(t^{\prime}\right)\right)\right].
\end{align}  Applying the short scattering time approximation similar to that discussed after Eq.~(\ref{interm-fn}) with the aid of $\left\langle\tilde{u}_{i^{\prime}}\right\vert\left.\tilde{u}_{i}\right\rangle=\delta_{i^{\prime},i}$, we are led to Eq.~(\ref{non-eq-nad}) in the maintext.


\begin{thebibliography}{References}
\bibitem{note-1} One can let the vector potential be dependent on $\boldsymbol{x}_{c}$, namely $\boldsymbol{A}=\boldsymbol{A}\left(t,\boldsymbol{x}_{c}\right)$. Then the equation for $\boldsymbol{k}_{c}$ will be changed from Eq.~(\ref{force-nq-2}) to $\hbar\dot{\boldsymbol{k}}_{c}=(-e)\left(\boldsymbol{E}+\dot{\boldsymbol{x}}_{c}\times\boldsymbol{B}\right)
-\left\langle \chi\right\vert\partial\mathcal{H}^{c}/\partial\boldsymbol{x}_{c}\left\vert \chi\right\rangle$, where $\boldsymbol{B}=\boldsymbol{\nabla}\times\boldsymbol{A}$, while other equations remain the same.
\bibitem{note-2-1} By a different gauge choice of the instantaneous basis $\left\vert{u}_{n}\left(\boldsymbol{\lambda}\right)\right\rangle\rightarrow
    \left\vert{u}^{\prime}_{n}\left(\boldsymbol{\lambda}\right)\right\rangle
=e^{i\theta_{n}\left(\boldsymbol{\lambda}\right)}\left\vert{u}_{n}\left(\boldsymbol{\lambda}\right)\right\rangle$, the coherence phase $\beta_{n,m}\equiv\text{arg}\left(\eta_{n}^{*}\eta_{m}/\left\vert\eta_{n}^{*}\eta_{m}\right\vert\right)$ is also changed $\beta_{n,m}\rightarrow\beta^{\prime}_{n,m}=\beta_{n,m}+\theta_{n}-\theta_{m}$, where $\boldsymbol{\lambda}$ denotes the centre-of-mass $(\boldsymbol{x},\hbar\boldsymbol{k})$. However, with Eq.~(\ref{ginvinsbasis}), the gauge change $\left\vert\bar{u}_{n}\left(\boldsymbol{\lambda}\right)\right\rangle\rightarrow
    \left\vert\bar{u}^{\prime}_{n}\left(\boldsymbol{\lambda}\right)\right\rangle
=e^{i\theta_{n}\left(\boldsymbol{\lambda}\right)}\left\vert\bar{u}_{n}\left(\boldsymbol{\lambda}\right)\right\rangle$ does not alter the value of $\bar{\beta}_{n,m}\equiv\text{arg}\left(\bar{\eta}_{n}^{*}\bar{\eta}_{m}/\left\vert\bar{\eta}_{n}^{*}\bar{\eta}_{m}\right\vert\right)$.
\bibitem{note-3} Since the degenerate active bands are fully occupied over the whole BZ, the part of the velocity obtained by summing the first term of Eq.~(\ref{phsvelo-sep-2-x}) (intra-manifold) over all wavepackets does not contribute to the current. Only the second term of Eq.~(\ref{phsvelo-sep-2-x}), describing the effects from other manifolds, contributes to the current. $\left\langle\dot{\boldsymbol{x}}\right\rangle$ is now obtained by summing the second term of Eq.~(\ref{phsvelo-sep-2-x}) over all the $N_{a}$ wavepackets. Substituting such $\left\langle\dot{\boldsymbol{x}}\right\rangle$ into Eq.~(\ref{Bltzeq-3}) gives $J_{\alpha}^{}\left(\boldsymbol{x}\right)\!=
e\!\int\!\!\text{d}\boldsymbol{k}\sum_{j=1}^{N_{a}}\sum_{\beta}
\left[\left\langle\mathcal{F}^{k_{\alpha}x_{\beta}}\right\rangle_{a}\dot{x}_{\beta}
+\left\langle\mathcal{F}^{k_{\alpha}k_{\beta}}\right\rangle_{a}\dot{k}_{\beta}
\right]_{j}$, where $J_{\alpha}^{}$ is the $\alpha$th spatial component of $\boldsymbol{J}^{}$. To arrive Eq.~(\ref{nAB-anomcrnt-0}), $\left(\dot{\boldsymbol{x}}_{},\hbar\dot{\boldsymbol{k}}_{}\right)$ above is found by solving Eq.~(\ref{phsvelo-sep-2}) to the zeroth order of $\epsilon$, Eq.~(\ref{epsilon}), so to keep $\boldsymbol{J}^{}$ to the first order of $\epsilon$.
\end{thebibliography}
\end{document}